# A Bias-Variance Framework for Estimating Stochastic Speed-Density Relationship Considering Constrained Sample Size

**Protiva Sadhukhan Prova**
Lecturer
Department of Civil Engineering (CE)
Bangladesh University of Engineering and Technology (BUET)
Email: protivaprova@ce.buet.ac.bd

**Nazmul Haque, PhD**
Assistant Professor
Accident Research Institute (ARI)
Bangladesh University of Engineering and Technology (BUET)
Email: nhaque@ari.buet.ac.bd
ORCID: *https://orcid.org/0009-0006-4698-3366*

**Md Asif Raihan, PhD***
Professor
Accident Research Institute (ARI)
Bangladesh University of Engineering and Technology (BUET)
Email: raihan@ari.buet.ac.bd
ORCID: *https://orcid.org/0009-0005-1762-8528*

**Md. Hadiuzzaman, PhD**
Professor
Department of Civil Engineering (CE)
Bangladesh University of Engineering and Technology (BUET)
Email: *mhadiuzzaman@ce.buet.ac.bd*
ORCID: https://orcid.org/0009-0004-0717-3569

***Corresponding Author**

# A Bias-Variance Framework for Estimating Stochastic Speed-Density Relationships Considering Constrained Sample Size

## Abstract

Accurate calibration of traffic speed–density fundamental diagram models is essential for freeway traffic flow analysis, yet existing approaches estimate parameters from fixed datasets without examining how estimates evolve as data availability increases. This study develops a comprehensive Bayesian stochastic calibration framework integrating sequential Markov Chain Monte Carlo (MCMC) estimation via a Metropolis-Hastings within Gibbs sampling scheme with a novel Bias–Variance Cone analysis for asymptotic parameter identification. The framework is applied to 17 macroscopic speed–density models using 1 full year of 5-minute PeMS loop detector data from 3 California freeway stations. Sequential MCMC progressively increases the number of randomly sampled days from 1 to 365, enabling systematic evaluation of parameter stability and convergence as a function of data availability. Among seventeen models, nine achieved stable posterior convergence with physically interpretable estimates, while four exhibited catastrophic structural failures due to logarithmic singularities, unit incompatibilities, and initialization sensitivity. The Bias–Variance Cone framework, the first application of asymptotic bias–variance decomposition to fundamental diagram calibration, establishes that free-flow speed parameters are reliably identified within ~30 days, whereas density and shape parameters require 100–182 days, providing the first quantitative minimum data requirements in the literature. Collectively, the findings provide an empirical identifiability ranking of 17 fundamental diagram models, a validated minimum data sufficiency framework, and a replicable, automated platform for large-scale stochastic fundamental diagram calibration.

# 1 Introduction

Understanding traffic flow dynamics is fundamental to the design, operation, and management of efficient transportation systems. The fundamental diagram (FD) provides a mathematical representation of the relationships among traffic flow, speed, and density and is widely used in traffic flow analysis, traffic management, and transportation system modeling. Although all three pairwise relationships, flow–density, speed–density, and speed–flow, are important, the speed–density relationship is particularly relevant to the representation of traffic dynamics in high-order traffic flow models (Messmer and Papageorgiou 1990; Papageorgiou and Kotsialos 2002). Its complexity stems partly from the influence of driving behavior and vehicle interactions on the observed relationship between speed and traffic density. Reliable estimation of this relationship therefore requires empirical observations that adequately represent different traffic conditions, including free-flow and congested regimes (Knoop and Daamen 2017).

Since Greenshields (1935) introduced the linear speed–density relationship, numerous FD models have been proposed with different mathematical structures, including logarithmic, exponential, and polynomial formulations (Greenberg 1959; Underwood 1960; Pipes 1966; Drew 1968). These models characterize traffic dynamics using parameters estimated by fitting a specified functional relationship to observed traffic data, typically via least squares or other optimization procedures. The estimated parameters are subsequently interpreted in terms of traffic characteristics, such as free-flow speed, jam density, and critical density, depending on the model formulation. However, each parametric model imposes a particular functional structure on the relationship between traffic variables. Consequently, its ability to represent empirical traffic behavior depends on the suitability of the assumed mathematical form and the characteristics of the calibration data. This limitation is particularly relevant in heterogeneous traffic streams, where differences in vehicle dimensions, performance, and driving behavior contribute to variations in longitudinal following, lateral movement, and effective use of road space.

Empirical traffic data also exhibit substantial variability arising from differences in driver behavior, vehicle composition, lane-changing maneuvers, road geometry, and external disturbances (Haque et al. 2025). Such variability is reflected in the scattering of observations around the central relationships represented in empirical flow–density and speed–density diagrams. Although some of this scatter reflects genuine fluctuations in traffic behavior, other components may arise from measurement errors, temporal variation, and unobserved factors. A deterministic FD model can describe the systematic component of the traffic relationship, but a single fitted curve does not, by itself, characterize the uncertainty associated with the estimated model parameters. Moreover, the estimates obtained from a particular dataset may depend on the number and distribution of observations available for calibration. Thus, a model that adequately fits a given sample does not necessarily yield parameter estimates that remain stable as the sample size changes.

This issue motivates the use of stochastic approaches to FD parameter estimation. Bayesian inference, implemented via Markov Chain Monte Carlo (MCMC) sampling, provides a framework for estimating parameter distributions while accounting for uncertainty under a specified likelihood and prior. Unlike a single point estimate obtained through conventional optimization, Bayesian calibration can characterize the posterior distribution of each parameter and support the assessment of uncertainty in parameter estimation. Nevertheless, obtaining posterior distributions does not, on its own, establish how parameter estimates evolve as the calibration sample increases. In particular, the relationship between sample size, estimation bias, variability, and parameter convergence requires explicit investigation. This distinction is important when traffic observations are limited, because the reliability of parameter estimates derived from a short observation period cannot be established solely from the goodness of fit achieved on that sample. **Figure 1** shows the variability in speed distribution for different sample sizes.

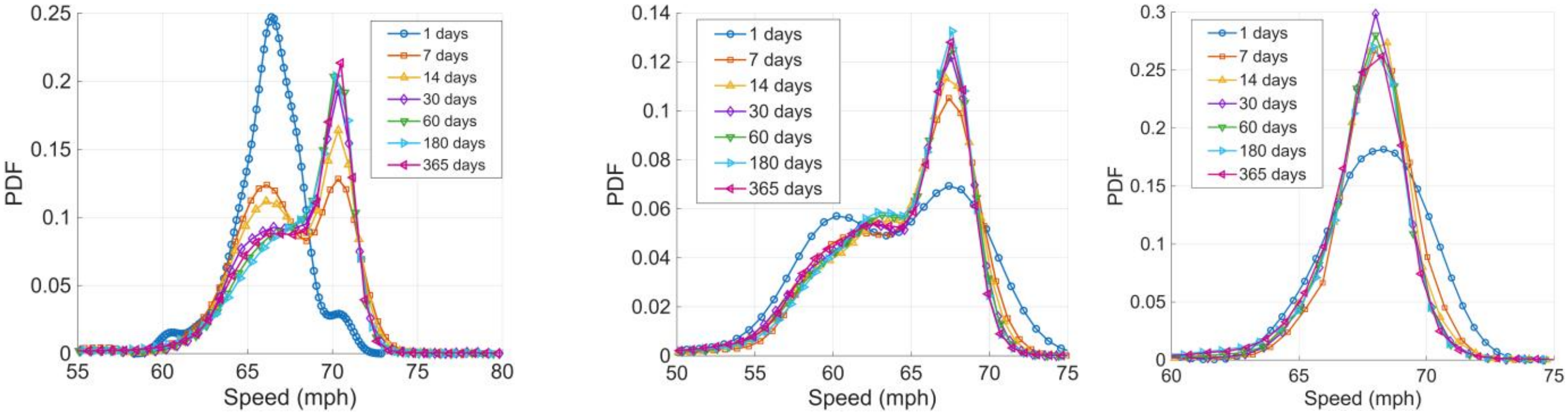

(a) 401892 (b) 401671 (c) 401657

**Figure 1.** Variability in speed distribution for different sample sizes (1 day to 365 days) (a) PeMS station 401892 (b) PeMS station 401671 (c) PeMS station 401657.

Despite the importance of sample size in statistical estimation, a systematic assessment of FD parameter convergence as sample size increases remains necessary to determine how reliably different model parameters approach their large-sample estimates. This challenge is especially relevant when comparing FD models with different functional structures and parameterizations, because their parameters may exhibit different convergence patterns. An explicit framework for examining the joint evolution of estimation bias and variability can therefore provide a more informative basis for assessing parameter stability than goodness-of-fit measures alone.

To address this issue, the present study investigates the sample-size-dependent convergence behavior of parameters in multiple speed–density FD models using Bayesian MCMC calibration and a Bias–Variance Cone framework. The framework is used to examine how parameter estimates and their associated bias–variance characteristics evolve as the calibration sample size increases toward the maximum considered. By comparing convergence patterns across models and parameters, the study seeks to identify differences in estimation stability and to assess the extent to which estimates from smaller samples approximate those from larger samples. The findings are intended to support more transparent evaluation of FD calibration reliability and to provide practical guidance for applications where empirical traffic observations are constrained.

# 2 Literature Review

## 2.1 The Fundamental Diagram and Speed–Density Models

The fundamental diagram (FD), describing the relationships among traffic flow, speed, and density, underpins applications ranging from shockwave propagation analysis (Lighthill and Whitham, 1955) to highway capacity and level-of-service determination (HCM 2022). Among the three pairwise relationships, the speed–density relationship is particularly central, as its calibration directly influences estimated traffic characteristics and the interpretation of traffic dynamics.

Since Greenshields (1935) introduced the linear speed–density model, numerous formulations have been developed encompassing power function (Pipes, 1966; Drew, 1968), logarithmic (Greenberg, 1959), exponential (Underwood, 1960; Newell, 1961; Castillo and Benítez, 1995), and logistic forms (Wang et al., 2011), valued for their mathematical tractability and interpretable parameters (Wu and Rakha, 2009). Subsequent developments introduced additional flexibility through the formulations of Papageorgiou et al. (1989), Kühne (1984), Lee et al. (1998), Gaddam and Rao (2019), Cheng et al. (2021), and Haque et al. (2025).

Despite differences in functional form, conventional parametric FD models assume a specified mathematical speed–density relationship, and their performance depends on the suitability of the selected form and the calibration data. A model may reproduce the dominant trend while exhibiting limitations near the free-flow to congestion transition (Edie, 1961; Drake, 1967). Multi-regime formulations partially address this but introduce additional challenges related to breakpoint identification and continuity at regime boundaries (Kidando et al., 2020). A further challenge is the substantial scatter commonly observed in empirical speed–density diagrams, arising from vehicle heterogeneity, driving behavior variation, lane-changing, road geometry, and measurement uncertainty, limitations that a single deterministic curve cannot fully characterize, motivating research into stochastic FD formulations and probabilistic calibration approaches (Kerner, 1998; Daganzo, 1994; Jabari and Liu, 2012; Qu et al., 2017).

To investigate these issues systematically, this study considers seventeen speed–density FD models spanning a range of established functional forms and parameter structures, summarized in **Table 1**, enabling examination of whether parameter convergence behavior varies across models of different mathematical complexity.

**Table 1:** Deterministic Speed–Density Fundamental Diagram Models

| Model | Mathematical Expression | Parameters ($\theta$) |
|---|---|---|
| Greenshields (1935) [GS] | $v = v_f\,(1 - \frac{k}{k_j})$ | $v_f,\ k_j$ |
| Greenberg (1959) [GB] | $v = v_c\ \log\left(\frac{k_j}{k}\right)$ | $v_c, k_j$ |
| Underwood (1960) [UW] | $v = v_f\ exp\left(-\frac{k}{k_c}\right)$ | $v_f, k_c$ |
| Northwestern (Drake et al., 1967) [NW] | $v = v_f\ exp\left[-\frac{1}{2}\left(\frac{k}{k_c}\right)^2\right]$ | $v_f, k_c$ |
| Pipes (1966) [PM] | $v = v_f\left(1 - \left(\frac{k}{k_j}\right)^n\right)$ | $v_f,\ k_j, n$ |
| Drake (1967) [DK] | $v = v_f\ exp\left[-\left(\frac{k}{k_c}\right)^2\right]$ | $v_f, k_c$ |
| Drew (1968) [DR] | $v = v_f\left[1 - \left(\frac{k}{k_j}\right)^{n+\frac{1}{2}}\right]$ | $v_f, k_j, n$ |
| Papageorgiou et al. (1989) [PG] | $v = v_f\ exp\left[-\frac{1}{\alpha}\left(\frac{k}{k_c}\right)^n\right]$ | $v_f, k_c, \alpha, n$ |
| Newell (1961) [NWL] | $v = v_f\left\{1 - exp\left[-\frac{\lambda}{v_f}(\frac{1}{k} - \frac{1}{k_j})\right]\right\}$ | $v_f, k_j, \lambda$ |
| Kühne (1984) [KH] | $v = v_f\left\{1 - \left(\frac{k}{k_j}\right)^{n_1}\right\}^{n_2}$ | $v_f, k_j, n_1, n_2$ |
| Kerner and Konhäuser (1994) [KN] | $v = v_f\left[\frac{1}{1 + exp\left(\frac{\frac{k}{k_c} - 0.25}{0.06}\right)} - 3.72 \times 10^{-6}\right]$ | $v_f, k_c$ |
| Castillo & Benítez, (1995) [CT] | $v = v_f\left\{1 - exp\left[\frac{\lvert w_j\rvert}{v_f}\left(1 - \frac{k_j}{k}\right)\right]\right\}$ | $v_f, k_j, w_j$ |
| Lee et al., (1998) [Lee] | $v = \frac{v_f\left(1 - \frac{k}{k_j}\right)}{1 + E\left(\frac{k}{k_j}\right)^n}$ | $v_f, k_j, E, n$ |

| Model | Mathematical Expression | Parameters ($\theta$) |
|---|---|---|
| Wang et al., (2011) [5PL] | $v = v_b + \frac{v_f - v_b}{\left\{1 + exp\left[\frac{k - k_t}{\theta_1}\right]\right\}^{\theta_2}}$ | $v_f, v_b, k_t, \theta_1, \theta_2$ |
| Gaddam & Rao, (2019) [GD] | $v = v_f \left(\frac{exp\left[-\left(\frac{k}{k_c}\right)^{1+n_1}\right] - exp\left[-\left(\frac{k_j}{k_c}\right)^{1+n_1}\right]}{1 - exp\left[-\left(\frac{k_j}{k_c}\right)^{1+n_1}\right]}\right)^{n_2}$ | $v_f, k_j, k_c, n_1, n_2$ |
| Cheng et al., (2021) [CH] | $v = \frac{v_f}{\left[1 + \left(\frac{k}{k_c}\right)^n\right]^{\frac{2}{n}}}$ | $v_f, k_c, n$ |
| Haque et al. (2025) [TAN] | $v = \tan\left[\tan^{-1}(v_f)\left(1 - \left(\frac{k}{k_j}\right)^n\right)\right]$ | $v_f, k_j, n$ |

## 2.2 Stochastic Fundamental Diagram Formulations

The limitations of deterministic FD representations have motivated substantial development in stochastic traffic flow theory. Kerner (1998) demonstrated experimentally, using three-phase traffic theory, that congested traffic occupies an extended two-dimensional region of the flow–density plane rather than a single curve, thereby redefining traffic breakdown as a stochastic phase transition. Daganzo (1994, 2006) showed, using the Cell Transmission Model, that empirical FD scatter can arise from stochastic variations in inflow and outflow conditions even under a simple underlying model. Fan and Seibold (2013) demonstrated that driver heterogeneity in free-flow speed produces a family of admissible traffic relationships rather than a single deterministic curve. Together, these contributions establish that the observed FD scatter arises from traffic-state non-uniqueness, demand stochasticity, and behavioral driver heterogeneity.

Building on these foundations, several studies developed probabilistic frameworks to model the distributional properties of the FD directly from empirical data. Jabari et al. (2012) analytically propagated Gamma-distributed microscopic driver heterogeneity to the macroscopic level, yielding percentile-based stochastic speed–density relations that bound empirical scatter. Ni et al. (2018) modeled speed as a Beta-distributed random variable conditioned on density, achieving R-squared of 0.9986. Qu et al. (2017) constructed percentile-based stochastic FDs without assuming any specific probability distribution using the theorem of total probability. Wang et al. (2013) combined a deterministic mean function with a Gaussian Process to capture density-dependent correlated stochasticity. Bai et al. (2021) showed that rainfall intensity and speed heterogeneity together explain 18-24% of the total FD variance. Siqueira et al. (2016) derived both the FD and its variance analytically from Stochastic Differential Equations within a Boltzmann-like framework. Kim et al. (2008) proposed a Switching Linear Dynamical System with variational Bayesian inference to model traffic state transitions at signalized intersections, defining the stochastic FD via the distribution of shockwave speeds.

For FD calibration, it is important to distinguish physical variability in observed traffic states, the primary concern of the stochastic formulations above, from inferential uncertainty in estimated model parameters arising from finite calibration datasets, which is the concern of Bayesian parameter estimation. Critically, none of the reviewed studies examine how FD parameter estimates evolve or converge as calibration data accumulates, leaving minimum data requirements for reliable parameter identification entirely unaddressed, the central motivation of the present study.

## 2.3 Calibration Methods and Bayesian Inference

FD calibration involves estimating model parameters from empirical traffic observations. Conventional least-squares (LS) estimation is widely used for its computational simplicity, but when free-flow observations substantially outnumber congested-state records, minimizing total squared error assigns disproportionate influence to the more frequently observed regime, producing calibrated relationships that fit free-flow conditions well but represent high-density conditions less accurately. Qu et al. (2015) investigated this issue and proposed a weighted least-squares (WLS) approach that assigns greater weight to underrepresented, high-density observations, demonstrating the importance of accounting for the observational distribution in FD calibration. Resampling approaches discussed by Zhang et al. (2018) provide a complementary strategy for addressing data imbalance. However, both optimization-based methods generally yield point estimates without quantifying uncertainty in the estimates.

Bayesian inference offers an alternative framework by representing unknown parameters as probability distributions that are updated based on observed data. In nonlinear FD calibration, Markov Chain Monte Carlo (MCMC) methods approximate posterior distributions when analytical solutions are unavailable. General treatments are provided by Gelman et al. (2013), with traffic-specific applications discussed by Muralidharan et al. (2011). Cheng et al. (2022) advanced FD calibration through a Bayesian Gaussian Process framework that models the residual covariance structure, connecting LS, WLS, and generalized least-squares formulations within a unified probabilistic representation that characterizes uncertainty beyond single-point estimates.

Nevertheless, Bayesian uncertainty quantification and parameter convergence with increasing sample size address fundamentally different questions. A posterior distribution characterizes uncertainty conditional on the available observations but does not establish how parameter estimates change as calibration data accumulates. Assessing convergence behavior requires repeated estimation under systematically varying sample sizes, distinguishing posterior uncertainty at each sample size from systematic differences between estimates obtained from smaller and larger datasets. This distinction directly motivates the sequential Bayesian MCMC framework developed in the present study.

### 2.4 Research Gap

The reliability of FD calibration depends not only on the selected model and estimation method but also on the amount and representativeness of available data. Yuan et al. (2010) examined systematic biases in loop-detector speed measurements and their implications for FD estimation, while Qu et al. (2015) and Cheng et al. (2022) investigated calibration approaches using fixed empirical datasets. Although these studies address important aspects of estimation and uncertainty, none examines how FD parameter estimates evolve as the calibration sample is systematically increased, a practically important question for studies relying on observations collected over limited periods.

Three questions motivate the present investigation: first, how do FD parameter estimates evolve as the calibration sample grows from a limited period toward a full year; second, how do finite-sample bias and estimation variability change across sample sizes and model structures; and third, how much data are required for parameter estimates to reach a stable level under the adopted estimation procedure.

To address this gap, the present study introduces a Bias–Variance Cone framework, combined with sequential Bayesian MCMC calibration, using 365 days of PeMS freeway data and examines parameter convergence behavior across seventeen speed–density FD models. The intended contribution is a systematic, model-comparative assessment of FD parameter convergence that can inform data-sufficiency decisions and improve interpretation of calibration results when empirical traffic observations are limited.

## 3 Methodology

### 3.1 Bias–Variance Cone Framework

#### 3.1.1 Theoretical Foundations

The Bias–Variance Cone framework draws on two established areas of statistical theory: the bias–variance decomposition of estimation error and asymptotic estimation theory, extending these concepts to analyze parameter convergence in traffic FD models.

For an estimator $\hat{\theta}_n$ of a target parameter $\theta$, the expected squared estimation error, expressed as the mean squared error (MSE), can be decomposed into squared bias and variance (Hastie et al., 2009):

$$\mathrm{MSE}(\hat{\theta}_n) = \mathbb{E}\left[\left(\hat{\theta}_n - \theta\right)^2\right] = \left(\mathbb{E}[\hat{\theta}_n] - \theta\right)^2 + \mathrm{Var}(\hat{\theta}_n) \tag{1}$$

Here, $\mathbb{E}[\hat{\theta}_n] - \theta$ represents the bias of the estimator, while $\mathrm{Var}(\hat{\theta}_n)$ represents its sampling variance. Both components may decrease as sample size increases, though their rates depend on the estimator, model specification, and sampling process. This decomposition provides a conceptual foundation for examining FD parameter convergence without implying monotonic decrease for every model or dataset.

Asymptotic estimation theory provides a basis for characterizing finite-sample bias behavior. Following Cox and Snell (1968) and Efron and Tibshirani (1993), the expected value of a regular estimator can be expressed as:

$$\mathbb{E}[\hat{\theta}_n] = \theta_\infty + \frac{b_1}{n} + \frac{b_2}{n^2} + \mathcal{O}(n^{-3}) \tag{2}$$

Where, $\theta_\infty$ denotes the limiting value of the estimator, and $b_1$ and $b_2$ are coefficients describing its leading finite-sample deviations. The applicability of this expansion to a particular FD parameter must be assessed in light of the estimator's properties and available data.

To accommodate potentially different empirical convergence patterns across FD models, variance decay is represented through a generalized power-law (van der Vaart, 1998):

$$\mathrm{Var}(\hat{\theta}_n) \approx \frac{\sigma^2}{n^\alpha} \tag{3}$$

Where $\sigma^2$ is a positive scale parameter and $\alpha$ is the variance-decay exponent. When $\alpha = 1$ , the expression corresponds to the conventional $1/n$ variance rate. Values of $\alpha > 1$ indicate faster variance decay, whereas values of $0 < \alpha < 1$ indicate slower decay relative to this benchmark. Slower decay may reflect weak parameter identifiability, observation dependence, or heterogeneous traffic conditions. Related methodological context is provided by Andrews (1993) on finite-sample bias in econometric estimation, Tierney and Kadane (1986) on Laplace approximations for Bayesian posterior quantities, and Severini (2000) on likelihood-based inference.

In the present study, these foundations motivate empirical investigation of sample-size-dependent parameter behavior across seventeen FD models using Bayesian MCMC estimation. An important distinction is that posterior uncertainty conditional on a given dataset is not equivalent to the variance of a repeated-sample estimator, and MCMC simulation error should not be conflated with either. Because the analysis covers a finite observation period, estimates from the largest available sample serve as maximum-sample reference values rather than true asymptotic values, providing a structured basis for assessing parameter stability, identifying differences in convergence across models, and evaluating the empirical adequacy of alternative bias and variance decay relationships.

### 3.1.2 Model Specification

By combining the inverse-sample-size bias expansion with a generalized variance-decay relationship, the Bias–Variance Cone framework represents the normalized parameter trajectory using a heteroscedastic Gaussian model:

$$\hat{\theta}_n \sim \mathcal{N}\left(\theta_\infty + \frac{b_1}{n} + \frac{b_2}{n^2}, \frac{\sigma^2}{n^\alpha}\right) \tag{4}$$

where the expected trajectory is defined as

$$\mu_n = \theta_\infty + \frac{b_1}{n} + \frac{b_2}{n^2} \tag{5}$$

Here, $\theta_\infty$ is the estimated limiting value of the normalized parameter trajectory, $b_1$ and $b_2$ characterize its leading finite-sample deviations, $\sigma^2$ is the variance-scale parameter, and $\alpha$ controls the rate at which the variance decreases with sample size. The mean trajectory approaches $\theta_\infty$ as $n \to \infty$ provided the bias terms decay as specified. Meanwhile, the standard deviation of the modeled deviations is $\sigma/n^{\alpha/2}$, Thus, the approximate 95% Gaussian uncertainty band around the mean trajectory is given by

$$\mu_n \pm 1.96 \frac{\sigma}{n^{\alpha/2}} \tag{6}$$

This band narrows as $n$ increases when $\alpha > 0$, producing the characteristic cone-like structure: greater modeled dispersion at smaller sample sizes and progressively smaller dispersion at larger sample sizes. The band represents the variability assumed by the cone model; it should not automatically be interpreted as a Bayesian credible interval or a confidence interval for the asymptotic parameter.

The framework takes as input the sequence of MCMC posterior mean estimates, $\{\hat{\theta}_1, \hat{\theta}_2, \ldots, \hat{\theta}_N\}$, where $\hat{\theta}_n$ denotes the posterior mean estimated from a sample containing $n$ randomly selected observation days, with $N$ as the total number of observation days (i.e., 365 days). To facilitate comparisons among parameters with different physical units and numerical magnitudes, each trajectory is normalized by its largest-sample estimate:

$$\tilde{\theta}_n = \frac{\hat{\theta}_n}{\hat{\theta}_N} \tag{7}$$

Where $\hat{\theta}_N$ is the posterior mean estimated using all $N$ days. This normalization sets $\tilde{\theta}_N = 1$ and expresses each estimate relative to its maximum-sample reference value. It facilitates comparisons of relative convergence patterns across parameters and models, although it does not eliminate differences in their convergence rates or guarantee that the fitted asymptotic limit equals unity. Normalization also requires $\hat{\theta}_N \neq 0$ ; interpretation may be problematic when the reference estimate is close to zero. The original parameter scale can be recovered by multiplying a normalized estimate by $\hat{\theta}_N$.

### 3.1.3 Parameter Estimation and Asymptotic Extrapolation

The five cone parameters $\{\theta_\infty, b_1, b_2, \sigma^2, \alpha\}$ are estimated by fitting the Gaussian cone model to the observed normalized parameter trajectory. Assuming a Gaussian working likelihood with variance $\sigma^2/n^\alpha$ , the negative log-likelihood is expressed as

$$\mathcal{L}(\theta_\infty, b_1, b_2, \sigma^2, \alpha) = \frac{1}{2}\sum_{n=1}^{N}\left[\log\left(\frac{2\pi\sigma^2}{n^\alpha}\right) + \frac{\left(\tilde{\theta}_n - \mu_n\right)^2}{\sigma^2/n^\alpha}\right] \tag{8}$$

Where, $\mu_n = \theta_\infty + b_1/n + b_2/n^2$ and $N = 365$. Minimizing this objective jointly estimates the asymptotic intercept, finite-sample bias coefficients, and variance-decay parameters. To enforce the positivity constraints on $\sigma^2$ and $\alpha$ , the optimization is conducted using unconstrained transformed parameters, with $\sigma^2 = \exp(\eta_\sigma)$ and $\alpha = \exp(\eta_\alpha)$ The resulting optimization problem can be solved using a quasi-Newton algorithm. Because normalized estimates across different sample sizes may be statistically dependent, this objective is interpreted as a Gaussian working likelihood, and the fitted cone parameters characterize the empirical convergence trajectory rather than automatically ensuring inferential validity.

Three complementary asymptotic estimates are computed to assess robustness. The model-based estimate is obtained directly from the fitted cone intercept:

$$\theta_\infty^{\text{model}} = \tilde{\theta}_\infty \cdot \hat{\theta}_N \quad \text{(direct cone intercept)} \tag{9}$$

Where, $\hat{\theta}_\infty$ is the fitted asymptotic intercept on the normalized scale. Multiplication by $\hat{\theta}_N$ converts this estimate back to the original parameter scale. The bias-corrected maximum-sample estimate adjusts the largest-sample normalized estimate for fitted finite-sample bias:

$$\hat{\theta}_\infty^{\text{last}} = \left(\tilde{\theta}_N - \frac{\hat{b}_1}{N} - \frac{\hat{b}_2}{N^2}\right)\hat{\theta}_N \tag{10}$$

Where $\hat{b}_1$ and $\hat{b}_2$ are the fitted bias coefficients. This estimate uses the largest available sample as the starting point and removes the bias predicted by the fitted cone model at $N$. The mean bias-corrected estimate averages bias-corrected values across all sample sizes:

$$\hat{\theta}_\infty^{\text{mean}} = \left[\frac{1}{N}\sum_{i=1}^{N}\left(\tilde{\theta}_i - \frac{\hat{b}_1}{i} - \frac{\hat{b}_2}{i^2}\right)\right]\hat{\theta}_N \tag{11}$$

Agreement among the three estimates indicates consistency under the fitted model but does not establish equality with the true population parameter. Finally, the fitted bias expansion can be used to extrapolate a parameter estimate from an observed sample size $n_s$ to a larger target sample size $n_L$ . On the normalized scale, the extrapolation is defined as

$$\tilde{\theta}_{n_L} = \tilde{\theta}_{n_s} + \hat{b}_1\left(\frac{1}{n_L} - \frac{1}{n_s}\right) + \hat{b}_2\left(\frac{1}{n_L^2} - \frac{1}{n_s^2}\right) \quad (12)$$

Converting the extrapolated value to the original parameter scale gives

$$\hat{\theta}_{n_L} = \hat{\theta}_{n_s} + \hat{\theta}_N\left[\hat{b}_1\left(\frac{1}{n_L} - \frac{1}{n_s}\right) + \hat{b}_2\left(\frac{1}{n_L^2} - \frac{1}{n_s^2}\right)\right] \quad (13)$$

Where $\hat{\theta}_{n_s}$ is the observed posterior mean at sample size $n_s$ and $n_L$ is the target sample-size horizon. The coefficients in these equations are estimated on the normalized scale, which is why the original-scale correction is multiplied by $\hat{\theta}_N$. As $n_L$ increases, the extrapolated estimate approaches the limit implied by the fitted bias expansion, provided the model remains appropriate beyond the observed sample-size range. Such extrapolation is therefore conditional on the adequacy of the fitted cone model and should be interpreted cautiously when the target horizon substantially exceeds the available 365-day observation period.

### 3.1.4 Performance Evaluation

The Bias–Variance Cone framework is evaluated in terms of extrapolation accuracy, goodness of fit, systematic error, and uncertainty calibration. For each sample size $i$, the extrapolation error is defined as:

$$e_i = \hat{\theta}_{L,i} - \tilde{\theta}_i \quad (14)$$

Where $\hat{\theta}_{L,i}$ denotes the estimate extrapolated from sample size $i$ to the target horizon $N$ and $\tilde{\theta}_i$ denotes the observed normalized parameter estimate at sample size $i$. The root mean squared error (RMSE), mean absolute error (MAE), and mean error (Bias) are used to quantify the magnitude and direction of the extrapolation errors:

$$\text{RMSE} = \sqrt{\frac{1}{N}\sum_{i=1}^{N} e_i^2} \quad (15)$$

$$\text{MAE} = \frac{1}{N}\sum_{i=1}^{N} |e_i| \quad (16)$$

$$\text{Bias} = \frac{1}{N}\sum_{i=1}^{N} e_i \quad (17)$$

RMSE penalizes larger errors more heavily, whereas MAE measures the average absolute deviation. Bias indicates the average direction of the errors: positive values indicate overestimation, while negative values indicate underestimation, according to the error definition above.

The goodness of fit of the estimated cone mean trajectory is evaluated using the coefficient of determination, $R^2$

$$R^2 = 1 - \frac{\sum_{i=1}^{N}\left(\tilde{\theta}_i - \hat{\theta}_{L,i}\right)^2}{\sum_{i=1}^{N}\left(\tilde{\theta}_i - \bar{\theta}\right)^2} \quad (18)$$

Higher $R^2$ values indicate that the fitted mean trajectory explains a greater proportion of the observed variation relative to a mean-only benchmark. However, $R^2$ should be interpreted alongside RMSE, MAE, and Bias because a high value does not necessarily imply accurate extrapolation or well-calibrated uncertainty bounds. Uncertainty calibration is assessed by examining the proportion of observed estimates that fall within the cone's predicted uncertainty bounds. Using an approximate two-standard-deviation interval, the empirical coverage is calculated as

$$\text{Coverage} = \frac{1}{N}\sum_{i=1}^{N} \mathbb{I}\left(\left|\tilde{\theta}_i - \hat{\mu}_i\right| \leq 2\hat{s}_i\right) \tag{19}$$

Where $\mathbb{I}(\cdot)$ is an indicator function equal to one when the condition is satisfied and zero otherwise, $\hat{\mu}_i$ is the fitted mean at sample size $i$ and $\hat{s}_i = \hat{\sigma}/i^{\hat{\alpha}/2}$ is the fitted standard deviation. Under a correctly specified Gaussian model, the interval $\hat{\mu}_i \pm 2\hat{s}_i$ corresponds approximately to 95% marginal coverage for individual observations around the modeled mean. Coverage evaluates whether observations fall within the uncertainty bounds around the fitted trajectory; a distinct quantity from extrapolation error; and should be interpreted as an in-sample diagnostic rather than an independent assessment of predictive calibration. All metrics are computed separately for each FD parameter and model, enabling identification of differences in convergence behavior, systematic errors, and adequacy of uncertainty characterization across the seventeen candidate formulations.

## 3.2 Sequential Bayesian MCMC Calibration Framework

### 3.2.1 Progressive Random Sampling Design

To systematically investigate the influence of data availability on fundamental diagram (FD) parameter estimation, a progressive random sampling design was implemented using daily traffic observations from the annual study period. At each iteration $i$ a subset of $i$ distinct days was randomly selected without replacement from the 365-day observation period. The corresponding calibration dataset was then constructed by pooling all eligible speed–density observations recorded on the selected days:

$$S_i = \{(k_t, v_t) : t \in D_i\} \tag{20}$$

Where $D_i$ denotes the set of selected observation days, and $k_t$ and $v_t$ represent the lane-normalized traffic density and corresponding speed observations, respectively. The sampling design ensures that the number of selected days increases progressively from $i = 1$ to $i = 365$, enabling parameter estimation to be evaluated across a broad range of data-availability levels. Each selected day contributes its available five-minute speed–density observations, approximately 288 intervals per complete day, so that the calibration dataset expands through the inclusion of complete daily records rather than arbitrarily selected isolated intervals. This design preserves within-day traffic condition sequences and incorporates cross-day variation into the calibration process, though it does not imply statistical independence among observations within or across days.

For each sample size, the dataset is used to calibrate the corresponding FD model via Bayesian MCMC, generating a sequence of parameter estimates and posterior distributions that characterize how parameter values, posterior uncertainty, and model fit evolve as additional daily observations become available. Applying this procedure across all 17 FD models and 3 detector stations yields 365 × 17 × 3 = 18,615 individual calibration runs. The resulting parameter trajectories serve as the primary empirical input for the Bias–Variance Cone analysis described in Section 3.1.

### 3.2.2 Fundamental Diagram Model Ensemble

Seventeen candidate FD models are evaluated as alternative functional representations of the equilibrium speed–density relationship, encompassing different mathematical formulations that enable examination of the sensitivity of parameter estimation to model structure within a common calibration framework. Each model is expressed in the general regression form:

$$y = f(x, \boldsymbol{\theta}) + \varepsilon \tag{21}$$

Where $x$ denotes traffic density, $y$ denotes traffic speed, $f(\cdot)$ represents the model-specific speed–density function, $\boldsymbol{\theta}$ is the corresponding parameter vector, and $\varepsilon$ represents the residual error associated with measurement noise and unexplained variation in the observed traffic states. The specific functional form and parameterization differ among the candidate models.

For each model, the calibration framework specifies four components: (i) the functional form $f(\cdot)$ defining the speed–density relationship; (ii) the model-specific parameter vector $\boldsymbol{\theta}$ , containing physically interpretable parameters and, where applicable, empirical shape coefficients; (iii) data-dependent initial parameter values $\boldsymbol{\theta}_0$ and (iv) prior distributions $p(\boldsymbol{\theta})$ that define the admissible parameter space and, where supported by prior knowledge, encode information about plausible parameter values. Depending on the model formulation, the parameters may include free-flow speed $\boldsymbol{v_f}$, jam density $\boldsymbol{k_j}$, critical density $\boldsymbol{k_c}$, and shape exponents such as $\mathbf{n}$, $\boldsymbol{n_1}$ and $\boldsymbol{n_2}$.

Initial parameter values are determined adaptively from the observations available in each sampled dataset rather than being held constant across all calibration iterations. This procedure accommodates changes in the observed traffic-state distribution as the number of sampled days increases. Specifically, free-flow speed $\boldsymbol{v_f}$ is initialized using the maximum observed speed in the sampled dataset, while jam density $\boldsymbol{k_j}$ is initialized using the maximum observed density. Critical density $\boldsymbol{k_c}$ is initialized as one-third of the maximum observed density, providing a simple data-dependent starting approximation to the density associated with capacity. Shape parameters $\mathbf{n}$, $\boldsymbol{n_1}$, and $\boldsymbol{n_2}$ are initialized at unity, and the residual standard deviation $\sigma$ is initialized at 0.2 mph. Adaptive initialization provides a consistent starting configuration across all 365 sample sizes but does not by itself guarantee physically plausible posterior estimates or prevent MCMC convergence problems, which must be assessed through prior specification and MCMC diagnostics.

The complete set of seventeen models, including mathematical expressions and parameter definitions, is presented in **Table 1**, enabling systematic comparison of how alternative speed–density formulations respond to progressively increasing data availability under the same sequential Bayesian calibration design.

### 3.2.3 Bayesian Inference Formulation

Parameter estimation is performed within a Bayesian framework, in which prior information about model parameters is updated using the observed speed–density data. According to Bayes' theorem, the posterior distribution of the parameter vector $\boldsymbol{\theta}$ conditional on the observed data $(\mathbf{x}, \mathbf{y})$ is given by

$$p(\boldsymbol{\theta} \mid \mathbf{x}, \mathbf{y}) = \frac{p(\mathbf{y} \mid \mathbf{x}, \boldsymbol{\theta}) p(\boldsymbol{\theta})}{p(\mathbf{y} \mid \mathbf{x})} \tag{22}$$

or, equivalently, in proportional form,

$$p(\boldsymbol{\theta} \mid \mathbf{x}, \mathbf{y}) \propto p(\mathbf{y} \mid \mathbf{x}, \boldsymbol{\theta}) p(\boldsymbol{\theta}) \tag{23}$$

Where $p(\mathbf{y} \mid \mathbf{x}, \boldsymbol{\theta})$ is the likelihood function, $p(\boldsymbol{\theta})$ denotes the prior distribution, and $p(\mathbf{y} \mid \mathbf{x})$ is the marginal likelihood, which serves as a normalizing constant. The posterior provides a probabilistic characterization of parameter uncertainty conditional on the observed data, specified likelihood, and adopted priors, and is used to examine parameter estimates and uncertainty across all 17 FD models and sample sizes. Each speed observation is modeled as:

$$y_i = f(x_i, \boldsymbol{\theta}) + \varepsilon_i, \qquad \varepsilon_i \sim \mathcal{N}(0, \sigma^2) \tag{24}$$

Where $x_i$ and $y_i$ denote the observed traffic density and speed, respectively; $f(x_i, \boldsymbol{\theta})$ is the model-specific FD function; and $\varepsilon_i$ is the residual error. The parameter $\sigma^2$ represents the residual variance, accounting collectively for measurement error and traffic-state variability not explained by the specified functional relationship. Under the assumption that residuals are conditionally independent given the model parameters, the likelihood for a dataset containing $N$ observations is

$$p(\mathbf{y} \mid \mathbf{x}, \boldsymbol{\theta}, \sigma^2) = \prod_{i=1}^{N} \frac{1}{\sqrt{2\pi\sigma^2}} \exp\left[-\frac{\left(y_i - f(x_i, \boldsymbol{\theta})\right)^2}{2\sigma^2}\right] \tag{25}$$

Here, $N$ denotes the number of speed–density observations in the calibration dataset, rather than the iteration index. The residual variance is estimated jointly with the structural parameters, allowing the model to characterize the dispersion of observed speeds around the fitted speed–density relationship.

Weakly informative normal priors are assigned to the structural parameters of all 17 FD models using a common prior specification:

$$\theta_j \sim \mathcal{N}(0, 10^2) \tag{26}$$

Where $\theta_j$ represents an individual structural parameter, including parameters associated with free-flow speed ($\boldsymbol{v_f}$), jam density ($\boldsymbol{k_j}$), critical density ($\boldsymbol{k_c}$) and model-specific shape or empirical coefficients such as $\mathbf{n}$, $\boldsymbol{E}$, $\boldsymbol{\lambda}$, $\boldsymbol{w_j}$, $\boldsymbol{v_b}$, $\boldsymbol{k_t}$, $\boldsymbol{\theta_1}$ and $\boldsymbol{\theta_2}$, where applicable. This common specification avoids model-specific prior tuning and establishes a consistent baseline for cross-model comparison, though the extent to which the likelihood dominates the posterior should be evaluated rather than assumed, particularly at small sample sizes. The residual variance is assigned an Inverse-Gamma prior:

$$\sigma^2 \sim \mathrm{IG}(\alpha = 2, \beta = 0.02) \tag{27}$$

This prior restricts the residual variance to positive values while regularizing its estimation. Together, the Gaussian likelihood and specified priors define a consistent inferential framework for investigating how parameter estimates and uncertainty respond to increasing data availability across all candidate FD models.

### 3.2.4 Metropolis-within-Gibbs MCMC Sampling

Because posterior distributions associated with nonlinear FD models lack closed-form solutions, parameter inference is performed using a Metropolis-within-Gibbs MCMC scheme. At each iteration, parameters are updated sequentially, conditioning each update on the current values of all remaining parameters, with each update performed via a Metropolis-Hastings step.

***Proposal mechanism***

For structural parameters updated directly in their natural parameter space, a Gaussian random-walk proposal is used. For parameter $p$, the candidate values at iteration. The candidate value at iteration $t+1$ is generated as

$$\theta_p{}' = \theta_p^{(t)} + \eta_p, \qquad \eta_p \sim \mathcal{N}(0, s_p^2) \tag{28}$$

Where $\theta_p^{(t)}$ is the current value of parameter $p$, $\theta_p{}'$ is its proposed value, and $s_p$ is the proposal standard deviation controlling the magnitude of the random-walk step. This proposal mechanism is used for structural parameters such as free-flow speed ($\boldsymbol{v_f}$), jam density ($\boldsymbol{k_j}$), critical density ($\boldsymbol{k_c}$), and model-specific shape coefficients, provided that their admissible parameter spaces permit unconstrained updates in the selected parameterization. For the residual standard deviation $\sigma$ , which must remain strictly positive, proposals are generated in logarithmic space:

$$\log(\sigma') = \log(\sigma^{(t)}) + \eta_\sigma, \qquad \eta_\sigma \sim \mathcal{N}(0, s_\sigma^2) \tag{29}$$

or, equivalently,

$$\sigma' = \sigma^{(t)} \exp(\eta_\sigma) \tag{30}$$

Since the Inverse-Gamma prior is defined for $\sigma^2$ , the likelihood and prior must be evaluated consistently using $\sigma^2 = (\sigma)^2$ .

***Acceptance criterion***

Each proposed update is evaluated using the Metropolis–Hastings acceptance probability. Let $\boldsymbol{\phi}$ denote the complete parameter vector, including the structural parameters and residual variance or standard deviation, as appropriate. For a proposal from the current state $\boldsymbol{\phi}^{(t)}$ to a candidate state $\boldsymbol{\phi}'$ the acceptance probability is

$$\alpha = \min\left[1, \frac{p(\mathbf{y} \mid \mathbf{x}, \boldsymbol{\phi}')p(\boldsymbol{\phi}')q(\boldsymbol{\phi}^{(t)} \mid \boldsymbol{\phi}')}{p(\mathbf{y} \mid \mathbf{x}, \boldsymbol{\phi}^{(t)})p(\boldsymbol{\phi}^{(t)})q(\boldsymbol{\phi}' \mid \boldsymbol{\phi}^{(t)})}\right] \tag{31}$$

Where $p(\mathbf{y} \mid \mathbf{x}, .)$and $q(\cdot|\cdot)$ denotes the likelihood density and the proposal density, respectively. For a symmetric Gaussian random-walk proposal in the parameter space in which the update is performed, the forward and reverse proposal densities cancel, yielding

$$\alpha = \min\left[1, \frac{p(\mathbf{y} \mid \mathbf{x}, \boldsymbol{\phi}')p(\boldsymbol{\phi}')}{p(\mathbf{y} \mid \mathbf{x}, \boldsymbol{\phi}^{(t)})p(\boldsymbol{\phi}^{(t)})}\right] \tag{32}$$

This simplified expression applies directly to symmetric proposals in the target parameter space. For proposals generated in transformed space, such as the log-space proposal for $\sigma$ , the corresponding proposal-density ratio must be included when evaluating the acceptance ratio on the original parameter scale, unless the target density has been correctly transformed to the proposal space.

To avoid numerical underflow when evaluating likelihoods involving many observations, the acceptance decision is computed on the log scale. For the general Metropolis–Hastings update,

$$
\begin{aligned}
\log r = {} & \log p\left(\mathbf{y} \mid \mathbf{x}, \boldsymbol{\phi}'\right) - \log p\left(\mathbf{y} \mid \mathbf{x}, \boldsymbol{\phi}^{(t)}\right) \\
& + \log p\left(\boldsymbol{\phi}'\right) - \log p\left(\boldsymbol{\phi}^{(t)}\right) \\
& + \log q\left(\boldsymbol{\phi}^{(t)} \mid \boldsymbol{\phi}'\right) - \log q\left(\boldsymbol{\phi}' \mid \boldsymbol{\phi}^{(t)}\right)
\end{aligned}
\tag{33}
$$

The candidate is accepted if

$$
\log u < \min(0, \log r), \qquad u \sim \text{Uniform}(0,1) \tag{34}
$$

and rejected otherwise, in which case the chain retains its current parameter value. Repeating these sequential updates generates a Markov chain whose stationary distribution is the specified joint posterior, provided that the transition kernel satisfies the relevant invariance and convergence conditions. Consequently, the resulting posterior samples can be used to estimate parameter summaries and quantify uncertainty in the calibrated FD models.

### 3.2.5 Multi-Chain Implementation and Adaptive Tuning

Two independent Markov chains are employed in each calibration run to facilitate posterior exploration and to assess sampling convergence. Both chains are initialized using the same starting parameter vector, $\boldsymbol{\theta}_0$, but evolve independently through separate stochastic proposal sequences. Where the MATLAB Parallel Computing Toolbox is available, chains are executed concurrently using a parallel for-loop with each chain assigned to a separate computational worker; otherwise, chains are executed sequentially using identical sampling configurations.

Each chain comprises 15,000 iterations, of which the first 5,000 are designated as burn-in and excluded from posterior inference. During this initial phase, the proposal distributions are adaptively tuned to improve sampling efficiency. After burn-in, every fifth iteration is retained, yielding 2,000 posterior samples per chain. The retained samples from the two chains are subsequently pooled to produce 4,000 posterior draws per parameter for each calibration run. The combined posterior sample matrix is expressed as

$$
\boldsymbol{\Theta} = \boldsymbol{\Theta}^{(1)} \cup \boldsymbol{\Theta}^{(2)} \tag{35}
$$

Where $\boldsymbol{\Theta}^{(1)}$ and $\boldsymbol{\Theta}^{(2)}$ denote the retained post-burn-in samples from the first and second chains, respectively. These pooled samples are used to calculate posterior summaries and quantify parameter uncertainty. Pooling is performed after burn-in and thinning; it does not, by itself, establish that the chains have converged or that the retained draws are effectively independent.

***Adaptive proposal tuning***

During burn-in, the proposal standard deviation $s_p$ for each parameter $p$ is adjusted at 500-iteration intervals according to the empirical acceptance rate. The update rule is

$$
s_p^{\text{new}} = \begin{cases} 1.1 s_p, & \text{if } a_p > 0.45, \\ 0.9 s_p, & \text{if } a_p < 0.15, \\ s_p, & \text{otherwise,} \end{cases} \tag{36}
$$

subject to the lower bound $s_p \geq 10^{-6}$, Where $a_p$ denotes the empirical acceptance rate for updates to the parameter p over the tuning interval. The rule increases the proposal scale when acceptance is high, decreases it when acceptance is low, and leaves it unchanged when acceptance falls within the target interval of 0.15–0.45. The initial proposal standard deviations are set to $s_p = 0.1$ for all structural parameters and $s_\sigma = 0.02$ for the residual standard deviation $\sigma$. The target acceptance interval of 0.15 to 0.45 is intended to balance parameter-space exploration against proposal rejection frequency. Roberts et al. (1997) established an asymptotic optimal acceptance rate of approximately 0.234 for high-dimensional random-walk Metropolis settings, providing theoretical context for the selected interval.

Adaptive tuning is restricted to the burn-in phase. At iteration 5,000, proposal standard deviations are fixed for the remaining 10,000 iterations, supporting a time-homogeneous Markov transition kernel during posterior sampling. The resulting retained draws are used for posterior inference subject to satisfactory convergence and mixing diagnostics.

### 3.2.6 Posterior Summarization and Model Evaluation

For each calibration run, posterior summaries are computed from the 4,000 retained samples obtained by pooling the post-burn-in samples from the two MCMC chains. For each model parameter, the summaries include the posterior mean, posterior standard deviation, posterior median (50th percentile), and 95% credible interval defined by the 2.5th and 97.5th percentiles of the sampled posterior distribution. These statistics characterize the central tendency, dispersion, and uncertainty of the parameter estimates conditional on the observed data and the specified Bayesian model.

To provide an additional measure of parameter precision, a standardized ratio of the posterior mean to its posterior standard error is calculated as

$$t_\theta = \frac{\bar{\theta}}{s_\theta/\sqrt{M}} \tag{37}$$

Where $\bar{\theta}$ is the posterior sample mean, $s_\theta$ is the posterior sample standard deviation, and $M$ is the number of retained posterior draws. In the present implementation, $M = 4{,}000$ . This ratio measures the posterior mean relative to its Monte Carlo standard error under an independent-sample approximation. However, because MCMC draws are generally autocorrelated, $s_\theta/\sqrt{M}$ may underestimate the actual Monte Carlo standard error. Moreover, this statistic is not a conventional frequentist $t$-statistic and should not be interpreted directly as a test of statistical significance or as definitive evidence of parameter identifiability. Parameter identifiability should be evaluated using the full posterior distribution, along with credible intervals, parameter correlations, and convergence diagnostics.

***Point prediction and predictive accuracy***

Point predictions of traffic speed are obtained by evaluating the model-specific FD function at the posterior mean of the structural parameter vector. Let $\boldsymbol{\theta}$ denote the vector of structural parameters, excluding the residual standard deviation $\sigma$ . The posterior mean parameter vector is

$$\hat{\boldsymbol{\theta}} = \mathbb{E}[\boldsymbol{\theta} \mid \mathbf{x}, \mathbf{y}] \tag{38}$$

and the predicted speed for observation $i$ is

$$\hat{y}_i = f\left(x_i, \hat{\boldsymbol{\theta}}\right) \tag{39}$$

Predictive performance is evaluated using the root mean square error (RMSE) and the coefficient of determination ($R^2$):

$$\text{RMSE} = \sqrt{\frac{1}{N}\sum_{i=1}^{N}(y_i - \hat{y}_i)^2} \tag{40}$$

$$R^2 = 1 - \frac{\sum_{i=1}^{N}(y_i - \hat{y}_i)^2}{\sum_{i=1}^{N}(y_i - \bar{y})^2} \tag{41}$$

Where $y_i$ and $\hat{y}_i$ are the observed and predicted speeds, respectively; $\bar{y}$ is the mean observed speed; and $N$ is the number of speed–density observations in the calibration dataset for the corresponding sampling iteration.

### 3.2.7 Output Structure and Computational Implementation

At the completion of each calibration run, posterior summary statistics, including the posterior mean, standard deviation, median, 95% credible interval bounds, parameter precision statistic, RMSE, and R-squared,  are appended incrementally to a model-specific structured output file as the sampling dataset increases from 1 to 365 days. This incremental strategy preserves completed results in the event of computational interruption and eliminates the need to repeat previously completed runs.

For each FD model-detector station combination, the output file contains up to 365 records, each representing a sampling iteration with its associated posterior summaries and model-fit metrics. These files serve as the primary inputs for the Bias–Variance Cone convergence analysis described in Section 3.1, enabling examination of parameter-specific convergence trajectories as sample size increases and comparison across all 17 FD models and three detector stations.

The complete calibration framework is implemented in MATLAB, with all 17 FD model functions, parameter definitions, initialization procedures, and prior specifications organized in a dedicated model library function that is called programmatically for each run, ensuring consistency across all sampling iterations and stations. The complete experiment comprises 17 models, 365 sampling iterations, and 3 detector stations, yielding 18,615 individual calibration runs, each consisting of two MCMC chains of 15,000 iterations, for a total of 558,450,000 chain iterations. Where available, MATLAB's Parallel Computing Toolbox executes the two chains concurrently within each run, improving computational throughput while preserving the specified sampling configuration.

# 4 Data Collection and Processing

## 4.1 Dataset Overview

Traffic data for the primary analysis was obtained from the California Department of Transportation (Caltrans) Performance Measurement System (PeMS), a statewide infrastructure monitoring platform that continuously aggregates real-time measurements from inductive loop detectors embedded in freeway pavement. Data were extracted at 5-minute aggregation intervals, offering an appropriate balance between temporal resolution and data volume for FD calibration.

Three detector stations were selected to represent different roadway configurations and enable cross-station generalizability assessment: (1) 401892 (6-lane freeway mainline); (2) 401671 (3-lane freeway mainline); (3) 401657 (4-lane freeway mainline).

For each station, one complete year of observations spanning September 2015 to August 2016 was collected, yielding approximately 105,000 five-minute records per station. Each record contains time-stamped measurements of per-lane traffic flow (veh/5-min), average speed (mph), aggregate station flow, composite speed, number of active lane detectors, and a percentage-observed field indicating the proportion of each interval measured directly rather than imputed by the PeMS algorithm.

## 4.2 Data Processing

Flow Conversion. Raw PeMS flow values are reported in vehicles per 5-minute interval. Standard hourly flow rates consistent with FD conventions were obtained via:

$$q = 12 \times q_{5\text{min}} \tag{42}$$

Where $q_{5\text{min}}$ is the observed station flow (veh/5-min) and $q$ is the equivalent hourly flow rate (veh/h)

Traffic density was computed from the fundamental flow identity:

$$k = \frac{q}{v} \tag{43}$$

Where $k$ is traffic density (veh/mile) and $v$ is average speed (mph).

Since the three selected stations differ in the number of lanes, density values were normalized on a per-lane basis to ensure comparability of traffic states across stations:

$$k^* = \frac{k}{L} \tag{44}$$

Where $k^*$ is lane-normalized density (veh/mile/lane) and $L$ is the number of lanes at the detector station. This normalization eliminates the systematic influence of roadway width on measured density, enabling speed–density relationships from all three stations to be analyzed within a common framework.

To support the progressive random sampling design, a day-of-year index (1 to 365) was extracted from each observation timestamp. This index serves as the primary sampling unit in the sequential calibration experiment, ensuring that each iteration represents a coherent daily traffic pattern rather than an arbitrary collection of isolated 5-minute intervals.

Following these processing steps, each observation is represented as a speed–density pair $(k^*, v)$, where lane-normalized density serves as the independent variable and speed as the dependent variable.

The final processed dataset for each station comprises approximately 105,000 five-minute speed–density observations spanning the full annual cycle of traffic conditions from free-flow to congestion, constituting the primary input to the Bayesian calibration.

# 5 Results and Discussion

## 5.1 Bayesian MCMC Calibration: Model Performance Classification

The sequential Bayesian MCMC calibration framework was applied to all seventeen FD models using the 365-day PeMS dataset across three stations. **Figure 2** presents the $R^2$ distributions computed across all sampling iterations from 1 to 365 days for each model at each station, providing a comprehensive picture of both predictive accuracy and stability under varying data availability.

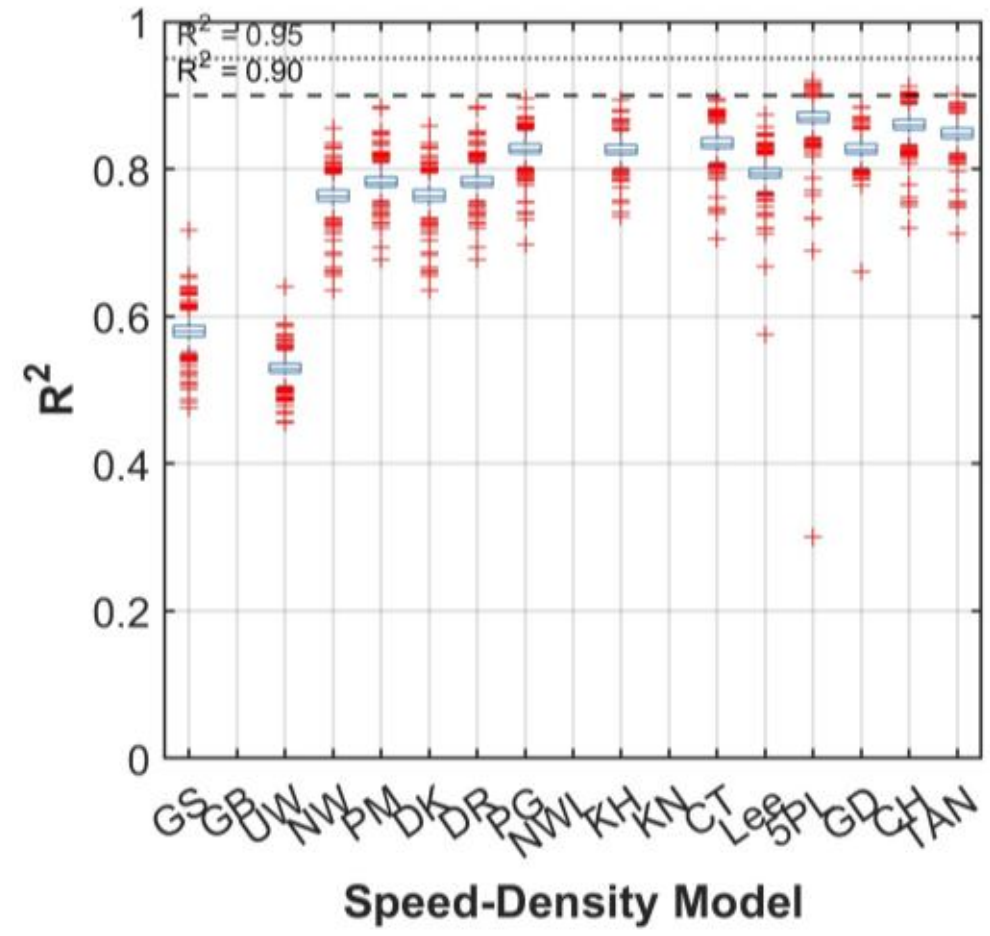

(a) Station 401892

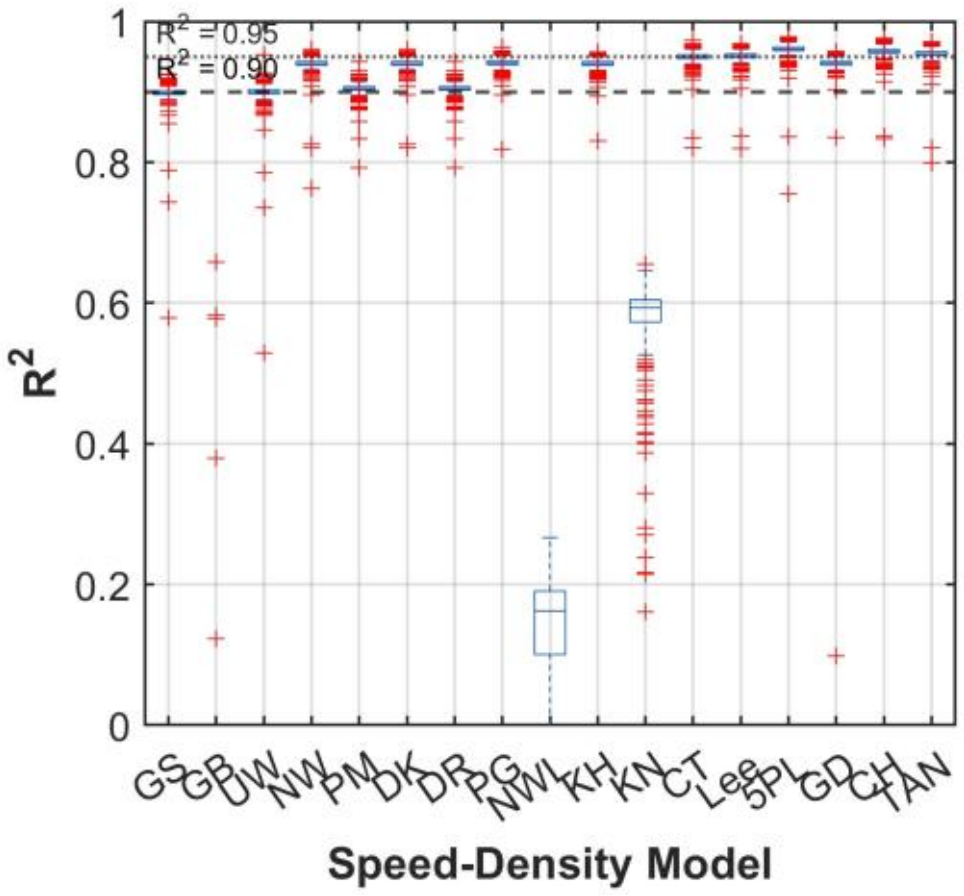

(b) Station 401671

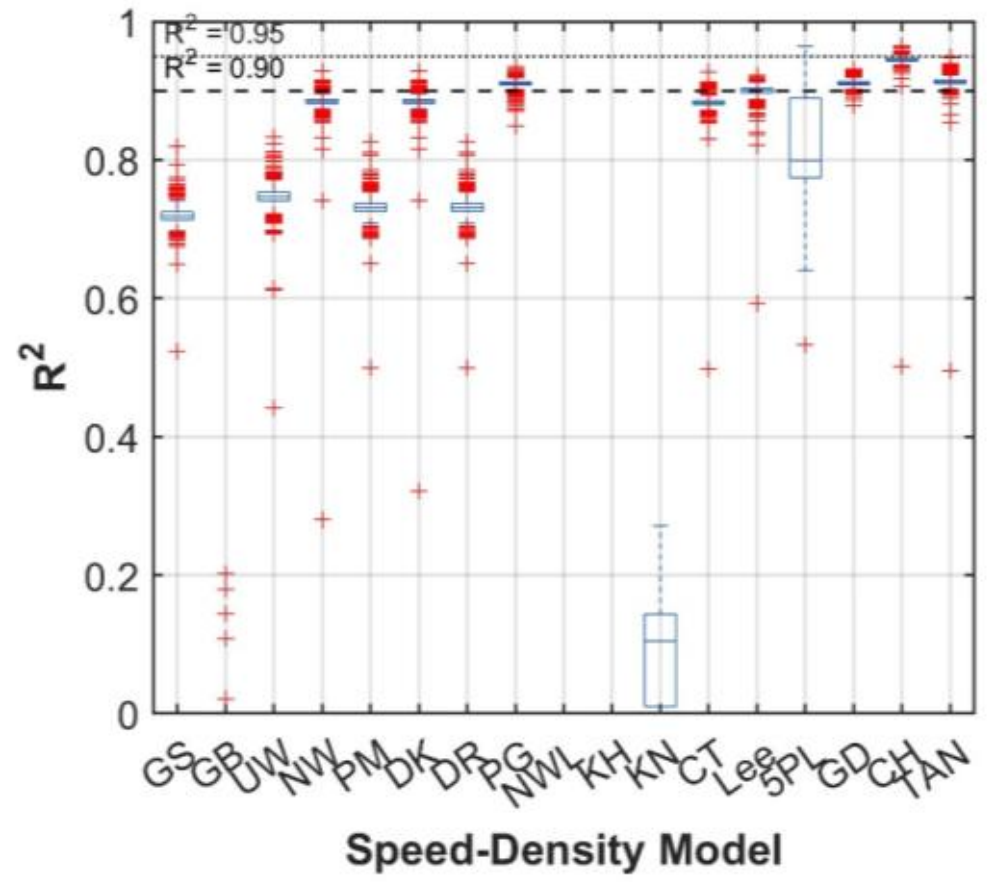

(c) Station 401657

**Figure 2.** Goodness of fit ($R^2$) values for different speed density models across different stations.

Across all three stations, a consistent performance hierarchy emerged among the structurally stable models. Station 401671 (3-lane) produced the highest overall $R^2$ values, followed by Station 401657 (4-lane), with Station 401892 (6-lane) yielding the lowest absolute $R^2$ levels, a pattern attributable to differences in lane configuration and traffic heterogeneity rather than model structure. At Station 401671, the CH model achieved the highest median $R^2$ of 0.957, followed by the TAN model (0.954), Lee (0.951), 5PL (0.950), CT (0.949), NW (0.940), and DK (0.940). At Station 401657, CH again led with a median $R^2$ of 0.944, followed by TAN (0.912), Lee (0.901), PG (0.910), NW and DK (both 0.884). At Station 401892, the 5PL model achieved the highest median $R^2$ of 0.869, followed by CH (0.859), TAN (0.847), CT (0.834), and Lee (0.794). The GS and UW models consistently produced the lowest median $R^2$ values among well-identified models across all stations, GS ranging from 0.579 at Station 401892 to 0.899 at Station 401671, and UW from 0.529 to 0.900, reflecting the well-known limitations of linear and simple exponential speed–density assumptions in capturing nonlinear freeway traffic behavior.

Critically, the standard deviations of $R^2$ across sampling iterations remained small for all well-identified models at all three stations. At Station 401671, standard deviations ranged from 0.010 for PM and DR to 0.023 for UW. At Station 401657, values ranged from 0.019 for PM and DR to 0.076 for 5PL. At Station 401892, values ranged from 0.017 for UW to 0.036 for 5PL. This cross-station consistency confirms that predictive performance is stable and not strongly sensitive to the specific random subset of days selected for calibration, implying that $R^2$ values achieved from as few as 30 to 50 days of data are reliable predictors of full-year performance, consistent with the convergence thresholds identified by the Bias–Variance Cone analysis.

The four structurally incompatible models produced catastrophically unstable $R^2$ distributions consistently across all three stations. The NWL model yielded entirely negative $R^2$ values at Station 401892 (mean of -36.27, minimum of -692.09) and Station 401657 (mean of -15.17, minimum of -5041.10), confirming complete structural incompatibility with PeMS 5-minute freeway data across all lane configurations. At Station 401671, NWL produced a median $R^2$ of 0.163, a mean of -1.28, and a minimum of -455.54, indicating episodic catastrophic failure even when occasional fits appear acceptable. The KN model similarly produced predominantly negative $R^2$ values across all stations, with means of -14.71, -0.96, and -8.27 at Stations 401657, 401671, and 401892, respectively, independently confirming unit incompatibility and hard-coded internal scaling failures. The GB model produced numerically corrupted $R^2$ values that reached the floating-point ceiling of 65,535 at all three stations, a direct consequence of the logarithmic singularity at low density, which generates unbounded speed predictions under free-flow conditions. The GD model exhibited similarly catastrophic overflow behavior, with mean $R^2$ values of $2.98 \times 10^{278}$, $3.47 \times 10^{34}, 4.24 \times 10^{221}$ at Stations 401657, 401671, and 401892 respectively, despite producing competitive median $R^2$ values of approximately 0.910, 0.940, and 0.826, confirming that GD is subject to catastrophic numerical failure events that make it entirely unreliable for systematic calibration regardless of station characteristics.

The KH model presents a particularly instructive case across all stations. At Station 401892, KH shows a competitive median $R^2$ of 0.825, but a mean of $1.43 \times 10^{31}$ and a maximum of $5.34 \times 10^{33}$. At Station 401671, a median of 0.940 is accompanied by a mean of $1.47 \times 10^{293}$ and a minimum of negative $-1.63 \times 10^{293}$. At Station 401657, the distribution collapses entirely with both mean and standard deviation reaching the floating-point ceiling of 65,535. This pattern demonstrates conclusively that median $R^2$ alone is an entirely misleading performance indicator for numerically unstable models, and that examination of the full $R^2$ distribution, including minimum values, means, and standard deviations, is essential for reliable model evaluation. Similarly, the PG model warrants caution: while achieving competitive medians of 0.910 and 0.941 at Stations 401657 and 401671, its standard deviations of 2.18 and 1.06, and minimum $R^2$ values of -41.06 and -19.49 at these stations reveal episodic convergence failures not evident from central tendency alone.

Taken together, the cross-station $R^2$ analysis provides strong empirical confirmation of the three-class model taxonomy established by the convergence diagnostics. The models CH, TAN, CT, Lee, NW, DK, PM, DR, and 5PL deliver stable and physically meaningful predictive accuracy across all three stations. The models GB, NWL, KN, and GD are structurally incompatible with high-frequency freeway loop detector data regardless of station characteristics. The models KH and PG exhibit partial reliability with episodic numerical instability that warrants caution in operational deployment. The consistent replication of this taxonomy across three stations with different lane configurations confirms the generalizability and robustness of the proposed Bayesian MCMC calibration framework.

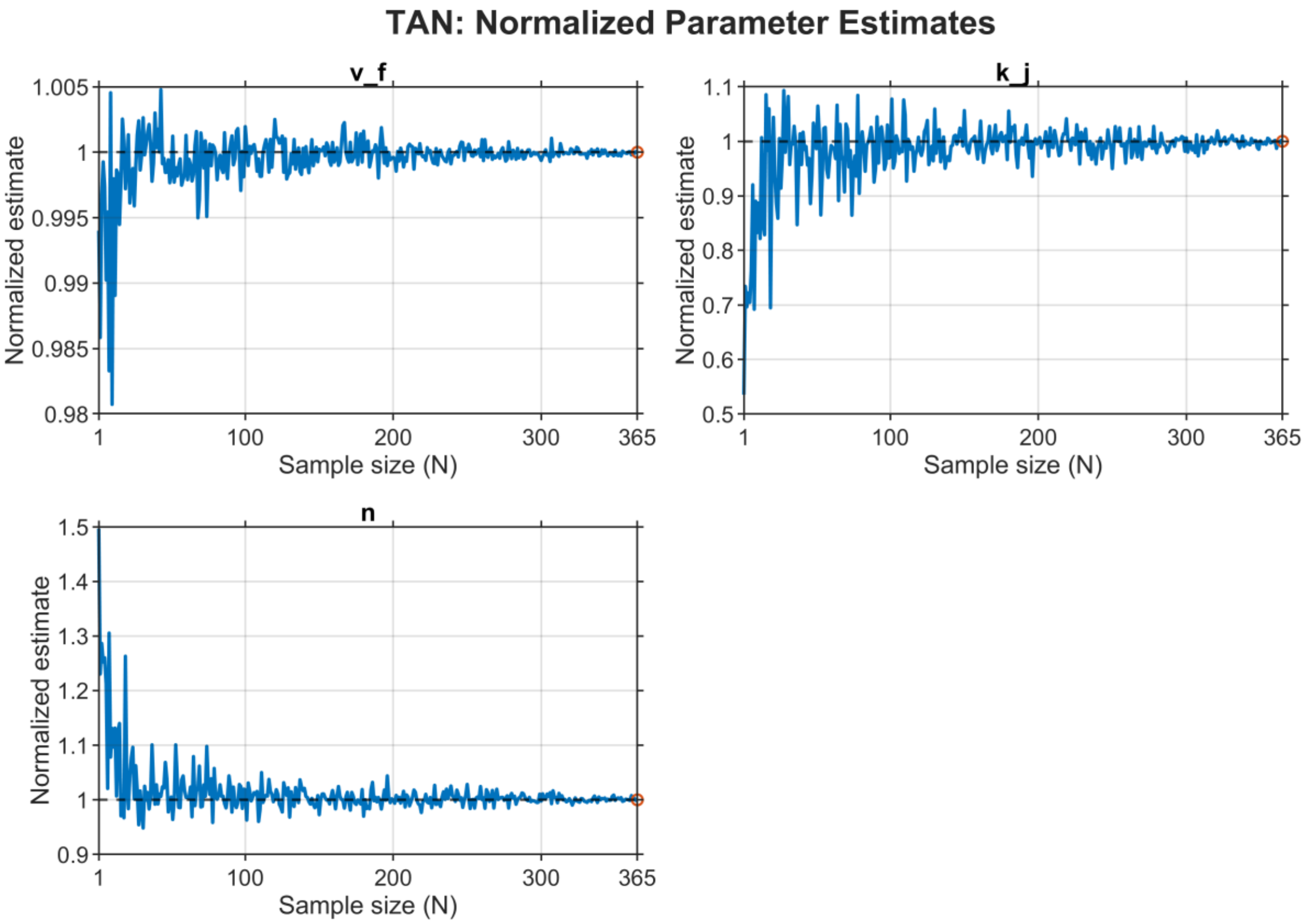
TAN: Normalized Parameter Estimates
v_f
k_j
n
Normalized estimate
Sample size (N)
1.005
1
0.995
0.99
0.985
0.98
1.1
0.9
0.8
0.7
0.6
0.5
1.5
1.4
1.3
1.2
1.1
0.9
1
100
200
300
365

(a)

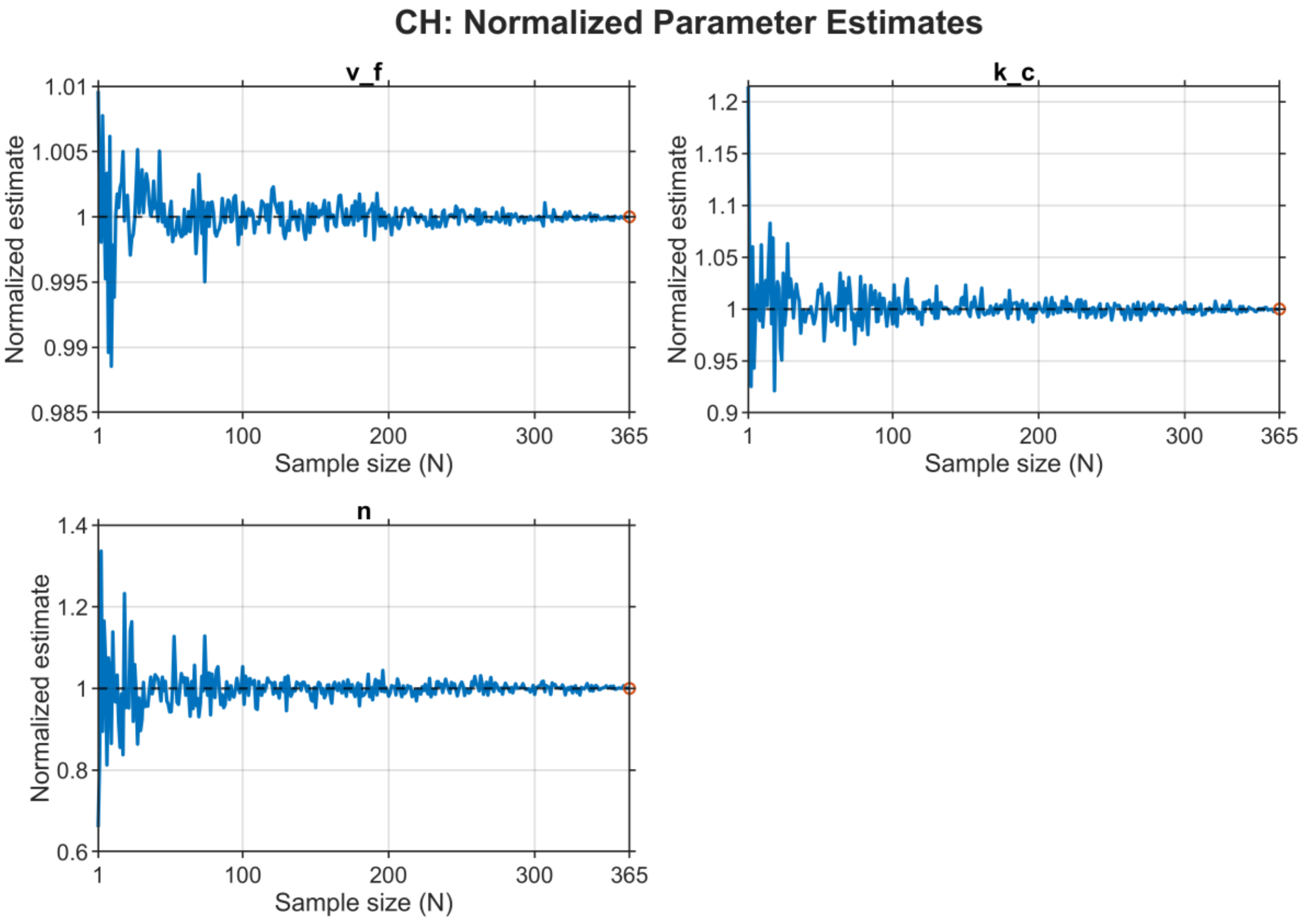
CH: Normalized Parameter Estimates
v_f
k_c
n
Normalized estimate
Sample size (N)
1.01
1.005
1
0.995
0.99
0.985
1.2
1.15
1.1
1.05
0.95
0.9
1.4
0.8
0.6
100
200
300
365

(b)

**Figure 3.** Illustration of the variation of the estimation Fundamental Diagram parameter values for different models (a) TAN; (b) CH (Station 401892)

**Figure 3** presents the sequential Bayesian MCMC parameter convergence trajectories for the TAN model (Haque et al., 2025) and the CH model (Cheng et al., 2021) across 365 progressive random sampling iterations. In each subplot, the blue line represents the normalized posterior mean estimate, the ratio of the posterior mean at iteration N to the full-sample estimate at N = 365, with the red circle at N = 365 denoting the reference convergence point and the dashed horizontal line at unity serving as the asymptotic baseline.

**Figure 3(a)** shows the parameter convergence of the TAN Model. The three parameters $\boldsymbol{v_f}$, $\boldsymbol{k_j}$, and $\mathbf{n}$ display clearly differentiated convergence behavior. Free-flow speed $\boldsymbol{v_f}$ oscillates within a narrow band of approximately 0.980 to 1.005 throughout the sampling range, with oscillation amplitude diminishing beyond $N$ = 100 days and stable convergence achieved by $n$ = 150 to 200 days, confirming reliable identifiability from as few as 30 to 50 days. Jam density $\boldsymbol{k_j}$ exhibits substantially greater early instability, dropping to approximately 0.50 at $n$ = 1 day and oscillating between 0.70 and 1.10 through N = 50 days, reflecting the physical scarcity of congested observations in short samples. Stable convergence is achieved by approximately $n$ = 150 to 180 days, consistent with Bias–Variance Cone findings that density parameters require 100 to 182 days for reliable identification. Shape exponent $\mathbf{n}$ shows intermediate behavior, beginning with deviations reaching approximately 1.50 at $n$ = 1 day before narrowing progressively, achieving stable convergence by $n$ = 150 to 200 days.

**Figure 3(b)** shows the parameter convergence of the CH Model. All three parameters exhibit broadly similar convergence patterns to the TAN model but with notably tighter oscillation bounds, reflecting the more parsimonious mathematical structure of the CH formulation. Free-flow speed $\boldsymbol{v_f}$ stabilizes within a narrow range of 0.985 to 1.010 and achieves convergence by $n$ = 50 to 100 days. Critical density $\boldsymbol{k_c}$ , governing density at maximum flow, is less volatile than $\boldsymbol{k_j}$ in the TAN model, converging stably by $n$ = 100 to 130 days. Shape exponent $\mathbf{n}$ spans approximately 0.60 to 1.40 at small sample sizes before narrowing to stable convergence by $n$ = 150 to 200 days.

Across both models, the consistent parameter identifiability hierarchy is confirmed: $v_f$ converges fastest, followed by density parameters, with shape exponents requiring the longest observation period, supporting the practical minimum data thresholds established by the Bias–Variance Cone framework.

## 5.2 Bias–Variance Cone Analysis

### 5.2.1 Asymptotic Parameter Convergence

The 365-day sequential MCMC parameter trajectories were analyzed using the Bias–Variance Cone framework to characterize asymptotic convergence behavior. The second-order inverse sample-size expansion:

$$\hat{\theta}_n = \theta_\infty + \frac{b_1}{n} + \frac{b_2}{n^2} \quad (45)$$

was fitted to each parameter trajectory, and the five cone parameters $\{\theta_\infty, b_1, b_2, \sigma^2, \alpha\}$. were estimated via negative log-likelihood minimization. The complete cone diagnostics for all seventeen models at Station 401892 are reported in **Table 2**.

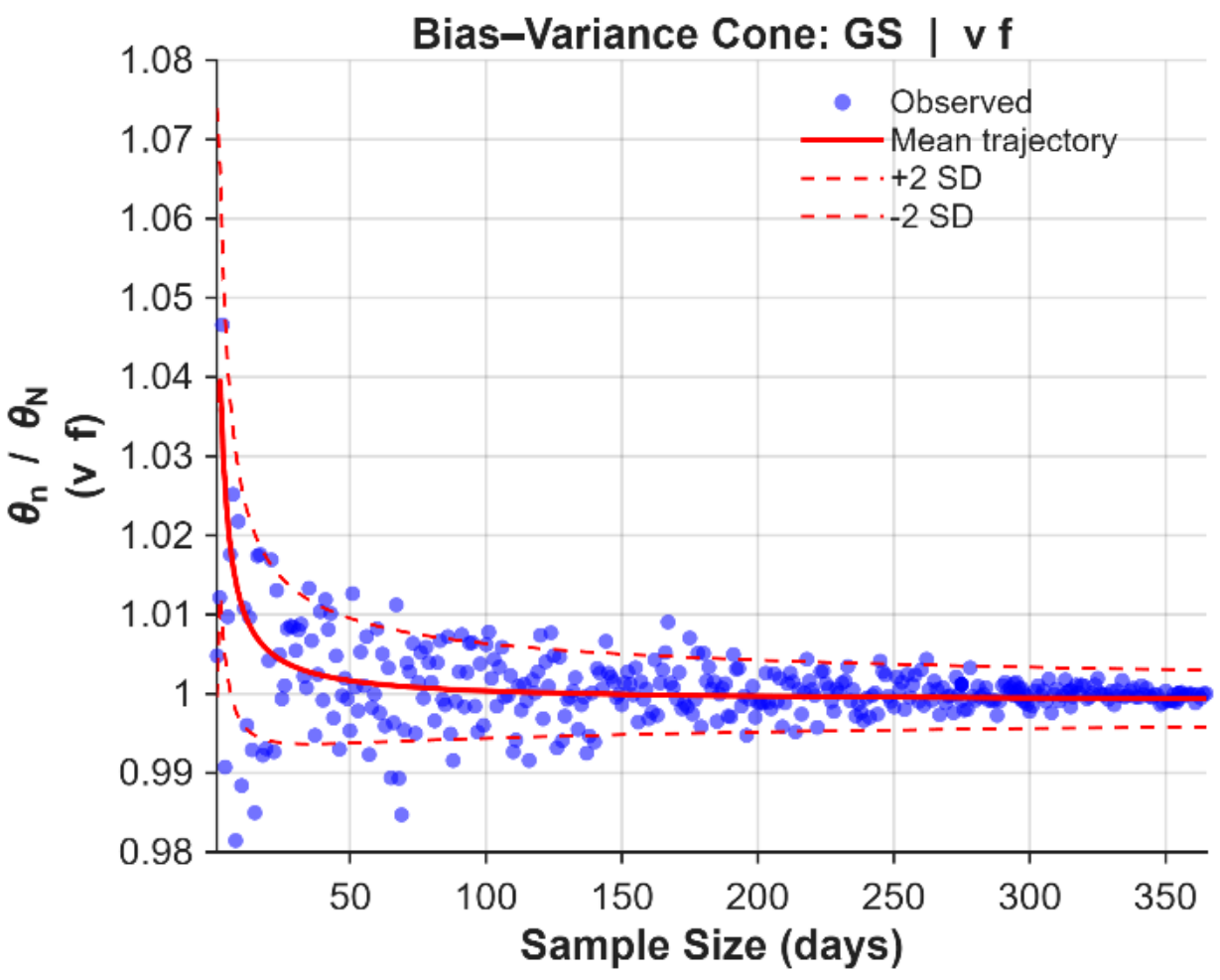
Bias–Variance Cone: GS | v f
Observed
Mean trajectory
+2 SD
-2 SD
$\theta_n$ / $\theta_N$ (v f)
Sample Size (days)

(a)

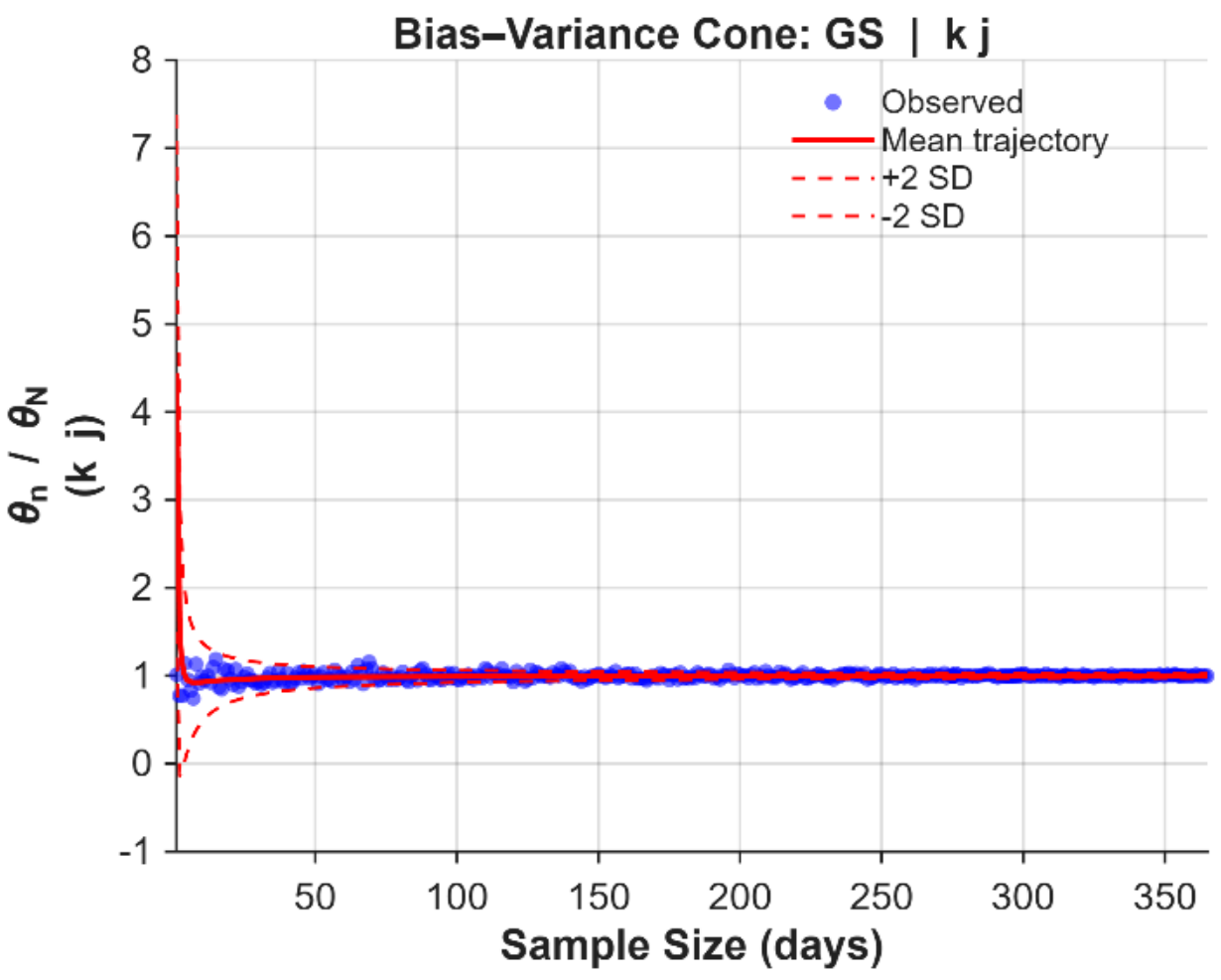
Bias–Variance Cone: GS | k j
Observed
Mean trajectory
+2 SD
-2 SD
$\theta_n$ / $\theta_N$ (k j)
Sample Size (days)

(b)

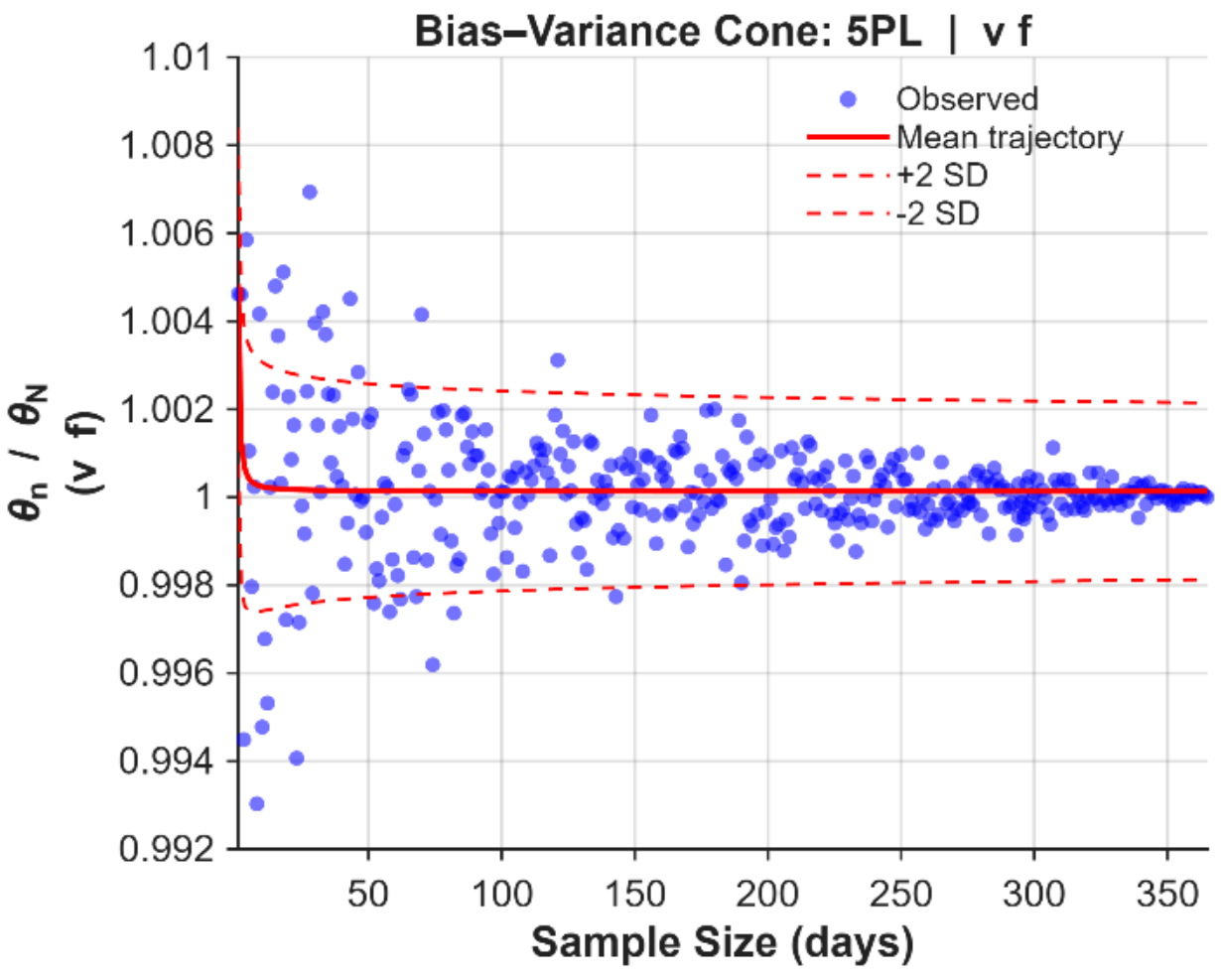
Bias–Variance Cone: 5PL | v f
Observed
Mean trajectory
+2 SD
-2 SD
$\theta_n$ / $\theta_N$ (v f)
Sample Size (days)

(c)

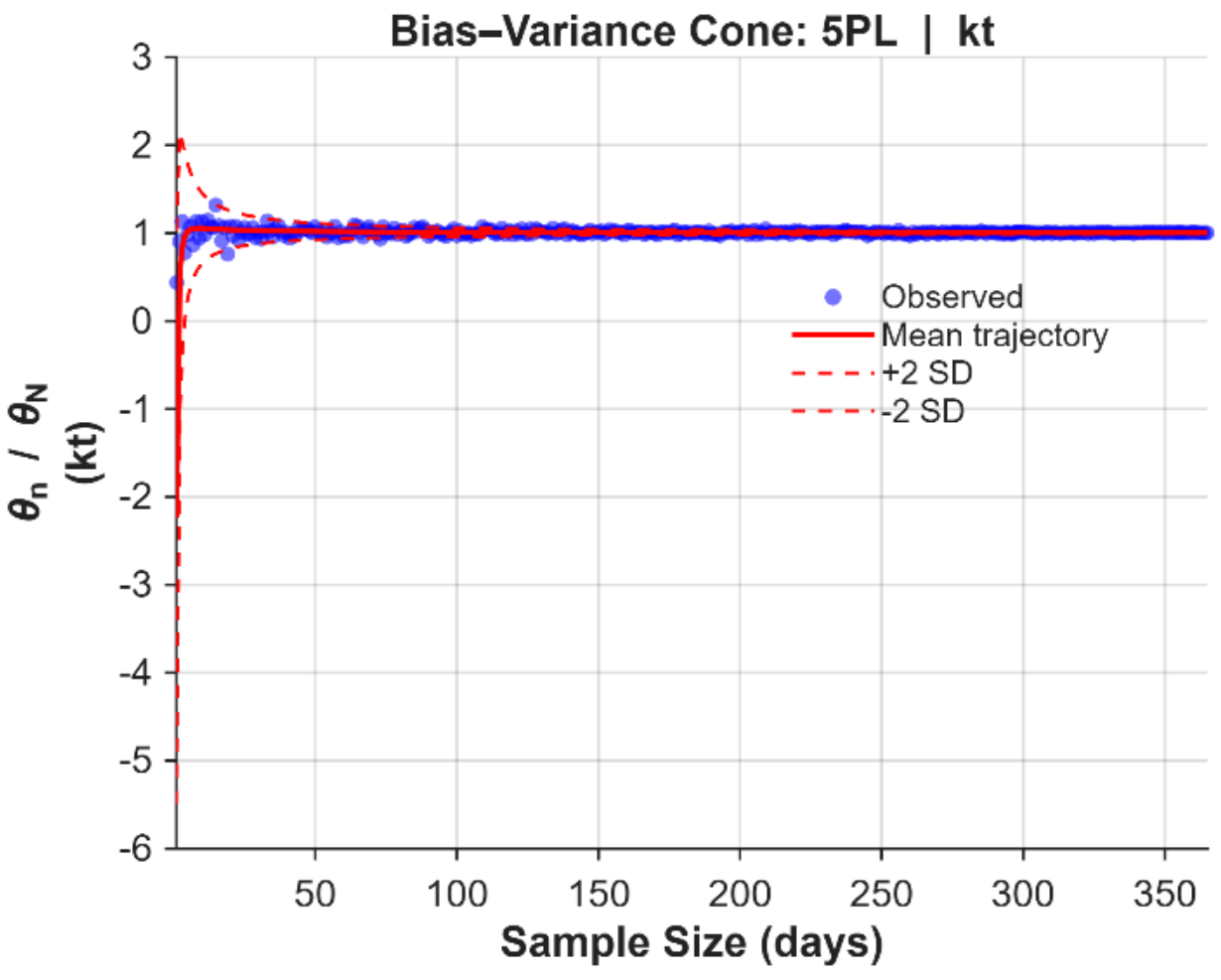
Bias–Variance Cone: 5PL | kt
Observed
Mean trajectory
+2 SD
-2 SD
$\theta_n$ / $\theta_N$ (kt)
Sample Size (days)

(d)

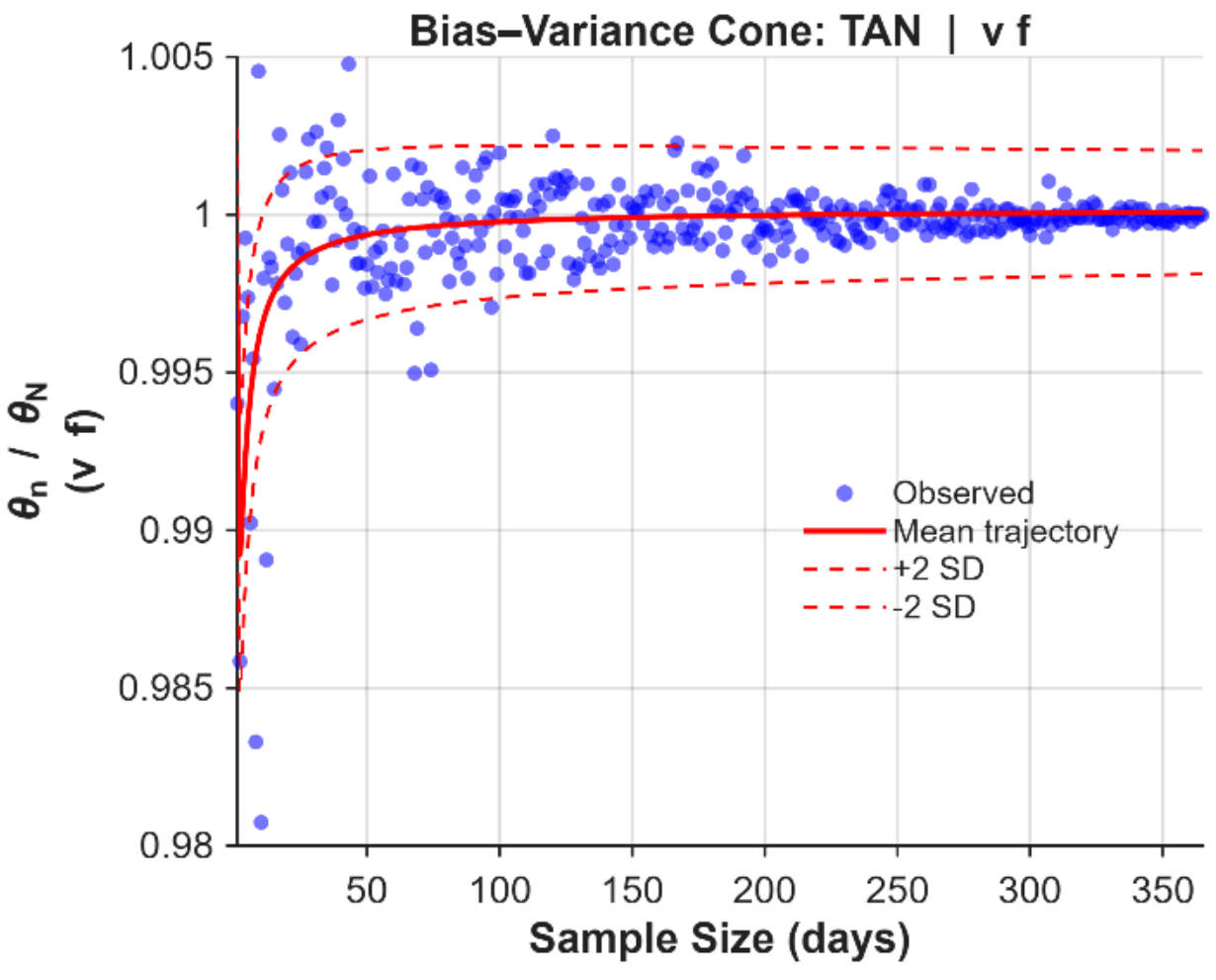
Bias–Variance Cone: TAN | v f
Observed
Mean trajectory
+2 SD
-2 SD
$\theta_n$ / $\theta_N$ (v f)
Sample Size (days)

(e)

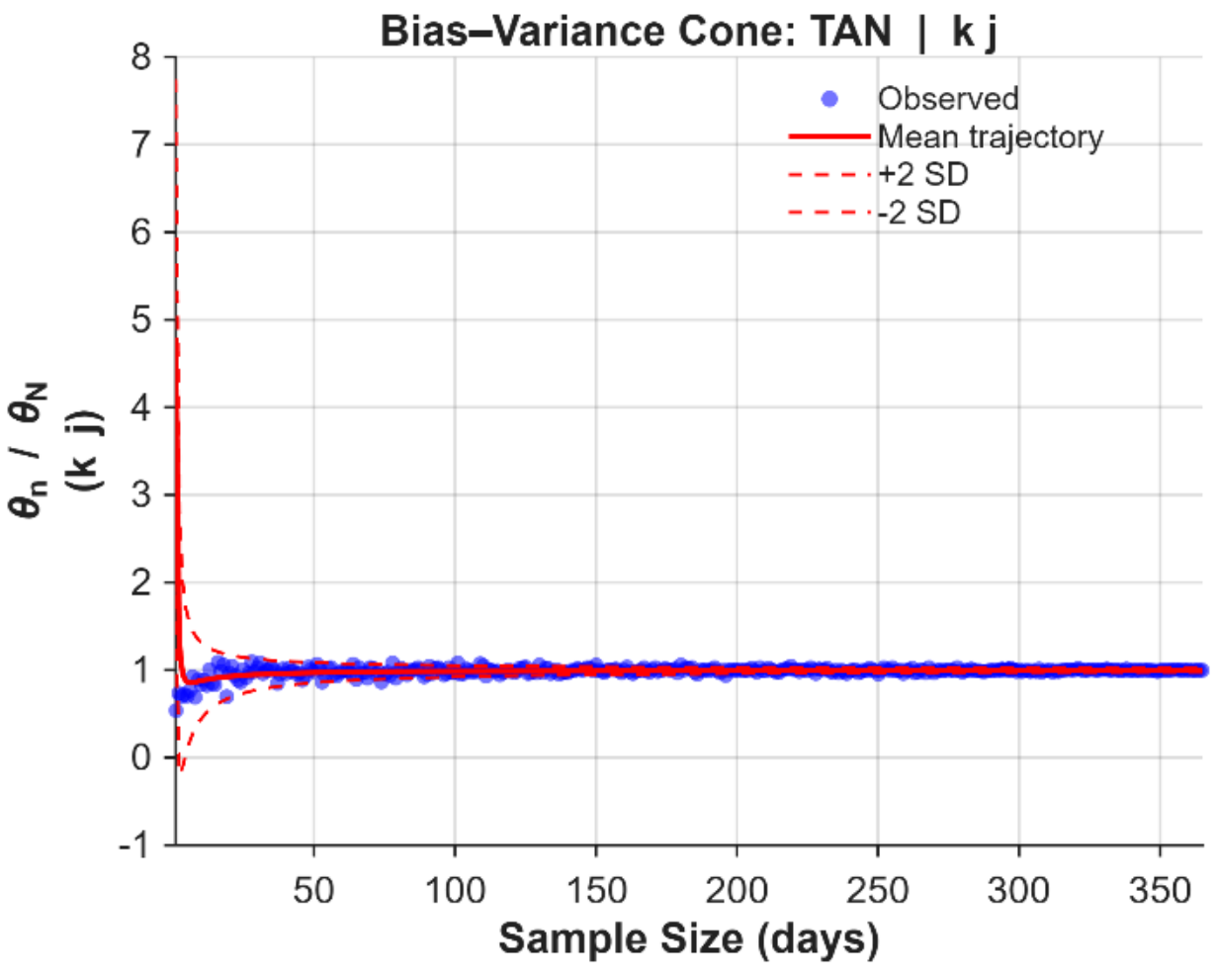
Bias–Variance Cone: TAN | k j
Observed
Mean trajectory
+2 SD
-2 SD
$\theta_n$ / $\theta_N$ (k j)
Sample Size (days)

(f)

**Figure 4.** Bias–Variance Cone Analysis for Selected Fundamental Diagram Models (Station 401892)

**Figure 4** presents Bias–Variance Cone trajectories for six representative parameter–model combinations drawn from the GS, 5PL, and TAN models, illustrating the convergence behavior of free-flow speed and density parameters across models of increasing structural complexity. In each panel , blue scatter points represent the normalized posterior mean estimates $\hat{\theta}_n/\hat{\theta}_N$ obtained from sequential Bayesian MCMC calibration at each sample size $n$ (1 to 365 days), the solid red curve represents the fitted mean trajectory $\mu_n = \theta_\infty + b_1/n + b_2/n^2$and the red dashed curves represent the $\pm$2SD uncertainty bounds of the Bias–Variance Cone. Convergence toward unity on the normalized scale corresponds to convergence of the posterior mean toward the asymptotic parameter value $\theta_\infty$.

The GS (**Figure 4(a)-(b)**) model represents the simplest two-parameter linear formulation and the archetypal well-identified model class. Free-flow speed $\boldsymbol{v_f}$ exhibits a monotonically decreasing trajectory from approximately 1.08 at $n$ = 1 day, settling toward unity by $n$ = 50 days, with cone bounds collapsing to approximately ±0.005 beyond $n$ = 150 days, confirming strong identifiability from moderate data volumes reflecting the physical dominance of free-flow observations in the PeMS dataset. Jam density $k_j$ presents markedly different behavior, with cone bounds spanning approximately 8 to −1 at n < 10 days before contracting sharply as congested observations accumulate, achieving stable convergence by n = 150 days and requiring approximately five times more data than $v_f$ for comparable posterior stability.

The 5PL model (**Figure 4(c)-(d)**), one of the most parameter-rich formulations in the ensemble with parameters $\boldsymbol{v_f}$, $\boldsymbol{v_b}$, $\boldsymbol{k_t}$, $\boldsymbol{\theta_1}$, and $\boldsymbol{\theta_2}$, illustrates partial identifiability behavior arising from over-parameterization. Free-flow speed $v_f$ converges within a tight normalized range of 0.992 to 1.010 by $n$ = 100 days, with uncertainty bounds of approximately ±0.004 at $n$ = 365 days, confirming strong identifiability despite model complexity. In contrast, inflection density $k_t$ exhibits dramatically wider cone bounds descending to approximately −6 at $n$ < 10 days. Although the mean trajectory stabilizes near unity by $n$ = 30 to 50 days, cone bounds remain wide throughout the annual range, confirming persistent estimation uncertainty due to limited congested-regime observations.

The TAN model (**Figure 4(e)-(f)**), a three-parameter trigonometric formulation with parameters $\boldsymbol{v_f}$, $\boldsymbol{k_j}$, and **n**, illustrates moderately well-identified behavior with a distinctive convergence pattern that contrasts with the GS model despite sharing the same parameter types. Free-flow speed $\boldsymbol{v_f}$ converges from below, beginning at approximately 0.982 at $n$ = 1 day and rising monotonically toward unity by $n$ = 100 days; reflecting the arctangent formulation's tendency to underestimate $\boldsymbol{v_f}$ from small samples, with uncertainty bounds retaining a perceptible width of approximately ±0.005 at $n$ = 365 days. Jam density $\boldsymbol{k_j}$ displays convergence qualitatively similar to GS $\boldsymbol{k_j}$, with extreme early dispersion followed by rapid contraction and stable convergence by $n$ = 100 days. The close similarity between GS and TAN $\boldsymbol{k_j}$ convergence behavior, despite their different functional forms, reinforces that $\boldsymbol{k_j}$ convergence rate is governed primarily by the frequency of congested observations rather than model-specific structural characteristics.

**Table 2** Bias–Variance Cone Analysis for 17 models for Station 401892 (365 Days Data)

| Model* | $\theta$ | $\theta_\infty^{model}$ | $\theta_\infty^{Last}$ | $\theta_\infty^{mean}$ | b1 | b2 | $\sigma^2$ | $\alpha$ | $R^2$ | RMSE | MAE |
|---|---|---|---|---|---|---|---|---|---|---|---|
| GS | $\boldsymbol{v_f}$ | 75.22 | 75.27 | 75.22 | 0.13 | -0.09 | 0.00 | 0.79 | 0.25 | 0.41 | 0.23 |
| GS | $\boldsymbol{k_j}$ | 205.30 | 205.06 | 203.38 | -1.23 | 4.67 | 2.16 | 1.63 | -15.96 | 37.82 | 7.45 |
| GB | $\boldsymbol{v_c}$ | 39.92 | 39.93 | 39.14 | -2.23 | 8.47 | 2.71 | 1.98 | -68.97 | 13.38 | 1.34 |
| GB | $\boldsymbol{k_j}$ | 41.70 | 42.96 | 42.44 | -7.31 | 17.17 | 23.58 | 1.59 | -0.82 | 25.17 | 5.31 |
| UW | $\boldsymbol{v_f}$ | 75.04 | 75.04 | 74.94 | 0.18 | -0.02 | 0.01 | 1.29 | -2.64 | 0.74 | 0.27 |
| UW | $\boldsymbol{k_c}$ | 194.94 | 194.75 | 193.53 | -1.56 | 4.52 | 1.98 | 1.57 | -8.68 | 31.42 | 7.31 |
| NW | $\boldsymbol{v_f}$ | 72.32 | 72.33 | 72.32 | 0.01 | -0.02 | 0.00 | 0.28 | 0.96 | 0.18 | 0.09 |
| NW | $\boldsymbol{k_c}$ | 131.13 | 131.03 | 131.24 | -0.27 | 0.48 | 0.00 | 0.43 | 0.56 | 2.47 | 1.25 |

| Model* | $\theta$ | $\theta_\infty^{model}$ | $\theta_\infty^{Last}$ | $\theta_\infty^{mean}$ | b1 | b2 | $\sigma^2$ | $\alpha$ | $R^2$ | RMSE | MAE |
|---|---|---|---|---|---|---|---|---|---|---|---|
| PM | $v_f$ | 71.41 | 71.40 | 71.40 | -0.02 | 0.01 | 0.00 | 0.99 | 0.86 | 0.17 | 0.09 |
| PM | $k_j$ | 102.38 | 102.44 | 100.38 | -0.63 | 5.14 | 2.62 | 1.95 | -34.79 | 26.53 | 3.36 |
| PM | n | 2.06 | 2.06 | 2.06 | 0.51 | 0.88 | 4.97 | 2.01 | -0.23 | 0.15 | 0.05 |
| DK | $v_f$ | 72.36 | 72.33 | 72.33 | 0.00 | -0.02 | 0.00 | 0.47 | 0.77 | 0.18 | 0.10 |
| DK | $k_c$ | 65.79 | 65.70 | 65.63 | -1.33 | 4.84 | 2.61 | 2.04 | -90.65 | 11.72 | 1.56 |
| DR | $v_f$ | 71.40 | 71.40 | 71.40 | -0.03 | 0.01 | 0.00 | 0.26 | 0.82 | 0.17 | 0.09 |
| DR | $k_j$ | 102.38 | 102.44 | 100.38 | -0.64 | 5.10 | 2.53 | 1.95 | -33.97 | 26.40 | 3.37 |
| DR | n | 1.56 | 1.56 | 1.56 | 0.67 | 1.30 | 8.54 | 2.00 | -0.39 | 0.16 | 0.05 |
| PG | $v_f$ | 70.75 | 70.75 | 70.75 | -0.02 | 0.01 | 0.00 | 0.21 | 0.92 | 0.13 | 0.07 |
| PG | $k_c$ | 46.19 | 45.66 | 46.18 | -5.32 | 6.07 | 0.56 | 0.99 | -1.38 | 5.14 | 2.84 |
| PG | a | 3.97 | 4.15 | 3.92 | 30.62 | -34.1 | 17.72 | 1.21 | -1.14 | 2.56 | 1.09 |
| PG | n | 2.96 | 2.96 | 3.00 | 0.56 | -3.87 | 2.02 | 1.72 | -15.37 | 0.54 | 0.10 |
| NWL | $v_f$ | 99.66 | 100.12 | 98.64 | 1.32 | -3.91 | 21.83 | 2.35 | -0.73 | 13.69 | 3.70 |
| NWL | $\lambda$ | 878.14 | 867.61 | 951.10 | -13.5 | -3.06 | 49.31 | 2.29 | -22.87 | 771.4 | 103.6 |
| NWL | $k_j$ | 552.33 | 564.02 | 898.87 | -60.5 | -15.1 | $1x10^5$ | 3.57 | -457.8 | 2221 | 355.5 |
| KH | $v_f$ | 70.73 | 70.74 | 71.66 | 0.05 | -2.80 | 10.00 | 3.18 | -27.48 | 10.67 | 0.97 |
| KH | $k_j$ | 253.65 | 256.60 | 310.54 | -21.5 | -7.90 | 607.5 | 2.60 | -179.5 | 416.6 | 62.13 |
| KH | $n_1$ | 2.96 | 2.95 | 2.77 | -0.72 | 21.64 | 1047 | 2.88 | -5.38 | 3.57 | 0.37 |
| KH | $n_2$ | 36.13 | 36.81 | 58.33 | -61.9 | -14.7 | $1x10^4$ | 2.86 | -374.4 | 146.9 | 23.65 |
| KN | $v_f$ | 819.20 | 822.08 | 839.02 | -0.23 | -13.9 | 20.08 | 2.35 | -23.39 | 581.0 | 59.33 |
| KN | $k_c$ | 587.94 | 582.56 | 929.90 | -55.4 | -16.3 | $3x10^7$ | 4.47 | -495.8 | 2194 | 351.4 |
| CT | $v_f$ | 70.28 | 70.28 | 70.28 | -0.01 | 0.01 | 0.00 | 0.22 | 0.98 | 0.12 | 0.06 |
| CT | $k_j$ | 114.68 | 114.78 | 111.91 | -0.67 | 6.76 | 3.22 | 1.89 | -81.04 | 37.82 | 4.70 |
| CT | $w_j$ | 72.46 | 72.30 | 72.55 | 0.54 | 0.08 | 5.35 | 1.82 | 0.73 | 5.69 | 2.28 |
| Lee | $v_f$ | 64.22 | 64.35 | 64.61 | 0.03 | -1.42 | 0.64 | 1.81 | -13.11 | 5.36 | 1.01 |
| Lee | $k_j$ | 103.98 | 104.01 | 103.93 | -0.16 | -0.94 | 1.21 | 1.81 | -0.37 | 4.88 | 1.74 |
| Lee | $E$ | -0.54 | -0.54 | -0.54 | 0.39 | 1.16 | 2.58 | 1.86 | -0.18 | 0.04 | 0.01 |
| Lee | n | 0.46 | 0.46 | 0.46 | 0.07 | 2.31 | 8.18 | 1.66 | 0.19 | 0.06 | 0.03 |
| 5PL | $v_f$ | 70.18 | 70.17 | 70.18 | 0.00 | 0.00 | 0.00 | 0.19 | 0.97 | 0.10 | 0.06 |
| 5PL | $v_b$ | 23.03 | 22.92 | 23.50 | 0.39 | -8.70 | 209.4 | 2.59 | -6.47 | 10.00 | 1.32 |
| 5PL | $k_t$ | 46.05 | 45.86 | 46.39 | 0.94 | -4.18 | 2.65 | 1.82 | -11.55 | 6.74 | 1.37 |
| 5PL | $\theta_1$ | 7.13 | 7.10 | 7.14 | 0.70 | -1.97 | 3.29 | 1.69 | -0.18 | 0.47 | 0.24 |
| 5PL | $\theta_2$ | 0.68 | 0.67 | 0.68 | 5.01 | -7.40 | 127.5 | 1.95 | 0.67 | 0.19 | 0.08 |
| GD | $v_f$ | 70.74 | 70.74 | 84.59 | 0.01 | 0.66 | $5x10^{12}$ | 7.93 | 1.00 | 70.62 | 14.16 |
| GD | $k_j$ | 119.00 | 117.98 | 124.93 | -8.60 | -0.87 | 149.9 | 2.38 | -9.28 | 60.10 | 10.60 |
| GD | $k_c$ | 50.21 | 49.04 | -32.60 | -20.0 | 509.3 | 7505 | 3.10 | -582.6 | 1226 | 95.88 |
| GD | $n_1$ | 1.98 | 1.98 | -1.55 | -3.72 | 455.5 | $8x10^5$ | 3.90 | -1239 | 47.78 | 3.77 |
| GD | $n_2$ | 0.27 | 0.26 | 0.27 | -18.5 | 80.48 | 1492 | 2.01 | -12.47 | 0.85 | 0.13 |
| CH | $v_f$ | 69.90 | 69.90 | 69.91 | -0.01 | 0.02 | 0.00 | 0.26 | 0.79 | 0.10 | 0.06 |
| CH | $k_c$ | 45.65 | 45.66 | 45.71 | 0.10 | -0.22 | 0.00 | 0.90 | 0.87 | 1.04 | 0.42 |
| CH | n | 5.21 | 5.21 | 5.26 | -0.02 | -2.67 | 2.31 | 1.78 | -9.56 | 0.70 | 0.16 |

| Model* | $\theta$ | $\theta_\infty^{model}$ | $\theta_\infty^{Last}$ | $\theta_\infty^{mean}$ | b1 | b2 | $\sigma^2$ | $\alpha$ | $R^2$ | RMSE | MAE |
|---|---|---|---|---|---|---|---|---|---|---|---|
| TAN | $v_f$ | 70.35 | 70.35 | 70.35 | -0.04 | 0.04 | 0.00 | 0.31 | 0.64 | 0.12 | 0.06 |
| TAN | $k_j$ | 218.40 | 218.52 | 215.57 | -1.83 | 5.62 | 2.17 | 1.66 | -11.35 | 49.87 | 8.36 |
| TAN | n | 3.81 | 3.81 | 3.91 | 1.44 | -7.67 | 3.64 | 1.95 | -44.56 | 1.38 | 0.16 |

*The abbreviations are mentioned in Table 1

**Table 2** summarizes the Bias–Variance Cone results for all 17 models at Station 401892, reporting three complementary asymptotic estimates ($\theta_\infty^{model}$, $\theta_\infty^{Last}$, $\theta_\infty^{mean}$), bias coefficients ($b_1$, $b_2$), variance parameters ($\sigma^2$, $\alpha$) and performance metrics ($R^2$, RMSE, MAE) for each parameter. The results reveal three distinct convergence classes. $\theta_\infty^{model}$ is the asymptotic parameter value extracted directly from the fitted cone intercept on the normalized scale and rescaled to original units; $\theta_\infty^{Last}$ the bias-corrected estimate derived from the largest available sample size N = 365 days by removing the finite-sample bias predicted by the fitted cone at that horizon; $\theta_\infty^{mean}$ the mean bias-corrected estimate obtained by averaging the bias-corrected normalized values across all 365 sampling iterations before rescaling, which aggregates convergence information from the entire trajectory rather than relying on any single sample size. The bias coefficients ($b_1$, $b_2$) describe the leading and second-order finite-sample deviations of the parameter trajectory from its asymptotic value, where large absolute values indicate slow convergence or persistent finite-sample bias, and opposing signs indicate non-monotonic or artificially compensating trajectory behavior. The variance parameters ($\sigma^2$, $\alpha$) characterize the stochastic uncertainty envelope of the cone: $\sigma^2$ is the variance scale controlling the overall width of the uncertainty bounds, and $\alpha$ is the variance decay exponent governing how rapidly the cone narrows as sample size increases, with $\alpha$ = 1 corresponding to the standard 1/n decay rate, $\alpha$ greater than 1 indicating faster-than-standard narrowing, and $\alpha$ less than 1 indicating slower narrowing reflecting weaker parameter identifiability.

Stably converging models, GS, UW, NW, PM, CT, and Lee, exhibit small bias coefficients, low variance, and consistent RMSE values, confirming well-identified parameter structures. Among these, GS and NW demonstrate particularly strong stability, with negligible long-run bias and low RMSE. Moderately converging models, PG, KH, and TAN, achieve acceptable predictive accuracy but retain persistent bias decay effects in shape parameters, indicating partial identifiability that persists even at large sample sizes.

Complete cone failures (RMSE > 100) occur exclusively in structurally incompatible models: NWL- $k_j$ (RMSE = 2221) KN-$k_c$ (2194) NWL-$\lambda$ (771), GD- $k_c$ (1226), and KH- $n_2$ (147). These extreme values indicate that the inverse-power bias structure cannot represent non-convergent or non-stationary parameter trajectories. In such cases, the optimization compensates by generating large bias coefficients of opposite signs, producing an artificial cancellation that improves the numerical fit only marginally while failing to capture the true trajectory dynamics. The consistently large negative $b_1$ values observed in these failures confirm that the optimization is attempting to fit upward or non-decaying trends that are structurally incompatible with the $1/n$ decay assumption of the cone model. Accordingly, high RMSE values should be interpreted not merely as poor numerical fits but as quantitative indicators of structural model incompatibility.

### 5.2.2 Effect of Sampling Duration on Asymptotic Estimates

To evaluate framework robustness under limited data availability, the Bias–Variance Cone analysis was repeated using randomly sampled datasets of 7, 30, and 182 days. Percentage deviations of asymptotic estimates $\theta_\infty^{model}$ from the 365-day benchmark are reported in Table 3.

Free-flow speed ($v_f$) is the most robustly identified parameter across all sampling durations. Deviations remained below 2% in most models even with 7-day datasets, confirming that free-flow characteristics can be reliably estimated from short observation periods.

The 182-day dataset produced estimates within 5% of the 365-day benchmark for most stable models (GS, UW, NW, PM, DK, DR, CT, CH, TAN, 5PL, Lee), suggesting that approximately six months of data is sufficient for reliable FD parameter identification in well-structured models. The 30-day dataset provided acceptable estimates for free-flow parameters but exhibited significant deviations in congestion-sensitive parameters, particularly $\boldsymbol{k_j}$, $\boldsymbol{k_c}$, and shape coefficients ($\mathbf{n}$, $\boldsymbol{n_1}$, $\boldsymbol{n_2}$, $\boldsymbol{\lambda}$), indicating that one month of data inadequately captures long-term traffic variability. The 7-day dataset produced the largest estimation errors across all parameter types and models, demonstrating that one week of data is generally insufficient for reliable asymptotic parameter estimation.

Structurally unstable models (GB, NWL, KH, KN, GD, PG) exhibited extreme sensitivity to reduced data lengths, with deviations exceeding 100–700% in several parameters, independently confirming their structural incompatibility with the Bayesian calibration framework.

These findings collectively establish that parameter convergence reliability depends not on data length alone but on whether the underlying parameter trajectory exhibits stable monotonic convergence toward its asymptotic value, a distinction the Bias–Variance Cone framework is uniquely positioned to diagnose.

### 5.2.3 Minimum Data Requirements for FD Parameter Identification

The Bias–Variance Cone analysis provides the first quantitative characterization of the minimum data requirements for reliable FD parameter identification across 17 models and 3 detector stations. **Table 3** reports the percentage deviation of the asymptotic estimate $\boldsymbol{\theta_\infty^{model}}$ from the 365-day benchmark $\boldsymbol{\theta_\infty^{Last}}$ at three data availability horizons, 7, 30, and 182 days, enabling direct cross-station comparison of parameter convergence sensitivity.

Free-flow speed ($\boldsymbol{v_f}$) is the most robustly identifiable parameter across all models and stations. For well-identified models at Station 401892, deviations at 182 days remain below 2%, GS $\boldsymbol{v_f}$ at 0.09%, NW $\boldsymbol{v_f}$ at 0.05%, CT $\boldsymbol{v_f}$ at 0.05%, and 5PL $\boldsymbol{v_f}$ at 0.02%, confirming reliable identification within approximately 30 days. This robustness is broadly replicated at Stations 401671 and 401657, where most well-identified models maintain $\boldsymbol{v_f}$ deviations below 5% even at 30 days, with the exception of PM and DR at Station 401671, which show elevated 182-day deviations of approximately 8%, attributable to station-specific congestion patterns rather than model structural limitations.

Density parameters ($\boldsymbol{k_j}$, $\boldsymbol{k_c}$) exhibit substantially greater sensitivity to sample size across all stations. At Station 401892, GS $\boldsymbol{k_j}$ shows 0.51% deviation at 182 days but 5.64% at 7 days. CT $\boldsymbol{k_j}$ maintains deviations below 5% at 182 days across all three stations, while NW $\boldsymbol{k_c}$ and CH $\boldsymbol{k_c}$ remain below 1% at 182 days at Station 401892, rising to approximately 10 to 25% at Station 401671 and Station 401657, respectively. These cross-station differences reflect the varying proportions of congested observations across lane configurations, with the 3-lane and 4-lane stations exhibiting greater sensitivity to the density parameter than the 6-lane station. Reliable density parameter identification generally requires 100 to 182 days across all well-identified models and stations.

Shape parameters ($\mathbf{n}$, $\boldsymbol{\theta_1}$, $\boldsymbol{\theta_2}$) exhibit the greatest sensitivity to sample size and the largest cross-station variability. PM $\mathbf{n}$ show deviations exceeding 100% at 182 days at Station 401657and DR $\mathbf{n}$ show deviations exceeding 100% at 182 days at Stations 401671 and 401657, while 5PL $\boldsymbol{\theta_2}$ reaches 178.82% at 7 days at Station 401892 and 102.14% at Station 401657. These results confirm that shape parameters generally require at least 182 days to achieve stable convergence in multi-parameter models, with convergence reliability varying substantially across stations.

Structurally incompatible models, GB, NWL, KH, KN, and GD, exhibit extreme deviations across all stations and horizons. GB $\boldsymbol{k_j}$ reaches 1131% at 182 days at Station 401657; GD $\boldsymbol{n_1}$ exceeds 15,667% at Station 401657, and KH $\boldsymbol{n_1}$ reaches 8,633% at 182 days at Station 401671, independently confirming their structural incompatibility with high-frequency loop detector data regardless of station characteristics or observation duration.

Collectively, these cross-station results establish three empirically validated minimum data thresholds: free-flow speed requires approximately 30 days; density parameters require 100 to 182 days; and shape parameters require 182 days or more. These thresholds are broadly consistent across the three stations despite differences in lane configurations and congestion regimes, confirming their practical applicability as general guidelines for assessing FD calibration data sufficiency in freeway monitoring contexts.

**Table 3** Percentage Deviation of Asymptotic Parameter Estimate $\boldsymbol{\theta}_{\infty}^{\boldsymbol{model}}$ from the 365-Day Benchmark ($\boldsymbol{\theta}_{\infty}^{\boldsymbol{Last}}$ ) Across Three Data Availability Horizons (7, 30, and 182 Days) for All Fundamental Diagram Models for Stations 401892, 401671 and 401657

| **Model*** | $\theta$ | **Station 401892** | | | **Station 401671** | | | **Station 401657** | | |
|---|---|---|---|---|---|---|---|---|---|---|
| | | **182** | **30** | **7** | **182** | **30** | **7** | **182** | **30** | **7** |
| GS | $v_f$ | 0.09 | 0.33 | 1.18 | 1.12 | 0.08 | 0.02 | 0.44 | 0.36 | 0.04 |
| GS | $k_j$ | 0.51 | 0.85 | 5.64 | 0.41 | 0.44 | 0.20 | 16.45 | 0.21 | 0.86 |
| GB | $v_c$ | 1.20 | 1.94 | 43.04 | 8.72 | 4.00 | 1.60 | 94.77 | 1.35 | 0.12 |
| GB | $k_j$ | 6.59 | 191.52 | 552.73 | 30.81 | 618.55 | 5.22 | 1131 | 193.82 | 267.29 |
| UW | $v_f$ | 0.18 | 0.46 | 1.89 | 4.65 | 0.11 | 0.11 | 2.79 | 0.22 | 0.03 |
| UW | $k_c$ | 0.21 | 1.90 | 10.09 | 36.12 | 0.55 | 0.87 | 39.23 | 2.00 | 0.56 |
| NW | $v_f$ | 0.05 | 0.18 | 0.63 | 1.29 | 0.05 | 0.06 | 1.34 | 0.14 | 0.01 |
| NW | $k_c$ | 0.20 | 0.96 | 2.95 | 10.39 | 0.19 | 0.20 | 24.84 | 1.28 | 0.27 |
| PM | $v_f$ | 0.04 | 0.11 | 0.20 | 8.07 | 0.40 | 0.32 | 1.67 | 0.31 | 0.04 |
| PM | $k_j$ | 3.05 | 4.01 | 0.93 | 19.75 | 0.26 | 0.07 | 9.62 | 3.21 | 0.04 |
| PM | n | 0.54 | 3.01 | 10.35 | 94.17 | 3.06 | 0.57 | 100.64 | 2.02 | 0.25 |
| DK | $v_f$ | 0.01 | 0.17 | 0.64 | 1.21 | 0.19 | 0.02 | 1.41 | 0.21 | 0.05 |
| DK | $k_c$ | 0.23 | 1.13 | 3.48 | 11.88 | 0.64 | 0.35 | 28.50 | 1.43 | 0.28 |
| DR | $v_f$ | 0.08 | 0.10 | 0.21 | 8.04 | 0.29 | 0.29 | 1.64 | 0.29 | 0.02 |
| DR | $k_j$ | 2.09 | 3.99 | 0.84 | 14.94 | 0.21 | 0.07 | 10.02 | 3.13 | 0.04 |
| DR | n | 0.66 | 3.90 | 13.80 | 236.21 | 7.61 | 1.40 | 248.22 | 5.11 | 0.59 |
| PG | $v_f$ | 0.02 | 0.18 | 0.29 | 5.22 | 0.30 | 0.28 | 1.75 | 0.44 | 0.02 |
| PG | $k_c$ | 6.26 | 20.95 | 31.92 | 76.11 | 41.73 | 17.93 | 43.51 | 29.70 | 2.69 |
| PG | $a$ | 12.59 | 110.69 | 327.27 | 1448.95 | 17.32 | 3.82 | -617.23 | -130.29 | -284.81 |
| PG | n | 1.96 | 4.40 | 3.00 | 53.13 | 2.12 | 0.40 | 22.90 | 2.32 | 0.14 |
| NWL | $v_f$ | 1.88 | 27.92 | 48.15 | 77.91 | 7.23 | 0.80 | 71.10 | 12.94 | 0.59 |
| NWL | $\lambda$ | 22.26 | 51.60 | 86.19 | 61.06 | 15.85 | 2.84 | 61.56 | 19.97 | 2.85 |
| NWL | $k_j$ | 21.18 | 79.29 | 91.84 | 87.18 | 75.75 | 23.62 | 88.13 | 78.74 | 30.94 |
| KH | $v_f$ | 1.75 | 65.14 | 0.94 | 112.50 | 3.11 | 2.05 | 2.46 | 5.34 | 0.74 |
| KH | $k_j$ | 21.21 | 44.09 | 58.33 | 99.49 | 73.21 | 47.44 | 15.85 | 10.29 | 0.03 |
| KH | $n_1$ | 3.22 | 766.34 | 1.53 | 8633.03 | 848.94 | 1035.75 | 51.43 | 9.57 | 1.93 |
| KH | $n_2$ | 45.68 | 86.10 | 93.66 | 85.07 | 80.91 | 58.94 | 5.76 | 12.97 | 4.84 |
| KN | $v_f$ | 19.11 | 10.52 | 79.30 | 66.88 | 12.19 | 5.45 | 66.95 | 11.00 | 5.10 |
| KN | $k_c$ | 31.71 | 65.50 | 90.81 | 46.87 | 26.37 | 26.59 | 75.81 | 53.08 | 42.86 |
| CT | $v_f$ | 0.05 | 0.15 | 0.01 | 2.62 | 0.26 | 0.08 | 1.31 | 0.01 | 0.03 |
| CT | $k_j$ | 2.85 | 4.45 | 5.05 | 20.04 | 0.69 | 0.02 | 9.81 | 1.55 | 0.03 |
| CT | $w_j$ | 1.04 | 2.13 | 6.86 | 57.09 | 1.89 | 0.15 | 64.86 | 3.80 | 0.06 |
| Lee | $v_f$ | 0.71 | 2.12 | 1.62 | 1.35 | 0.61 | 0.10 | 6.43 | 4.60 | 0.02 |
| Lee | $k_j$ | 0.56 | 3.42 | 6.19 | 75.62 | 57.40 | 22.87 | 45.51 | 29.43 | 16.86 |
| Lee | $E$ | 0.64 | 4.66 | 7.42 | 92.45 | 87.99 | 61.45 | 90.71 | 83.34 | 30.22 |
| Lee | n | 1.47 | 15.28 | 15.71 | 36.49 | 0.94 | 0.56 | 21.70 | 1.54 | 0.67 |
| 5PL | $v_f$ | 0.02 | 0.06 | 0.12 | 3.60 | 0.04 | 0.02 | 0.01 | 0.19 | 0.06 |

| Model* | $\theta$ | Station 401892 | | | Station 401671 | | | Station 401657 | | |
|---|---|---|---|---|---|---|---|---|---|---|
| | | 182 | 30 | 7 | 182 | 30 | 7 | 182 | 30 | 7 |
| 5PL | $v_b$ | 4.59 | 8.14 | 21.62 | 73.10 | 3.52 | 0.53 | 30.22 | 19.96 | 9.23 |
| 5PL | $k_t$ | 0.14 | 0.55 | 16.00 | 12.42 | 5.22 | 1.46 | 19.81 | 18.50 | 10.83 |
| 5PL | $\theta_1$ | 0.60 | 1.13 | 0.71 | 20.51 | 8.64 | 4.30 | 28.49 | 40.90 | 27.64 |
| 5PL | $\theta_2$ | 1.04 | 20.00 | 178.82 | 3.07 | 13.24 | 3.62 | 102.14 | 90.80 | 59.55 |
| GD | $v_f$ | 145.36 | 87.97 | 356.24 | 953.22 | 347.62 | 790.33 | 351.43 | 40.64 | 30.94 |
| GD | $k_j$ | 4.87 | 32.99 | 83.27 | 95.80 | 65.59 | 20.22 | 61.00 | 35.98 | 14.28 |
| GD | $k_c$ | 512.30 | 360.06 | 786.83 | -160.80 | -155.32 | -229.92 | 141.79 | 52.70 | 27.22 |
| GD | $n_1$ | 50.55 | 284.21 | 587.54 | 804.92 | 224.32 | 20.27 | 15667 | 2520 | 1245 |
| GD | $n_2$ | 37.11 | 23.66 | 291.19 | 17.69 | 48.46 | 49.74 | 35.47 | 56.97 | 31.31 |
| CH | $v_f$ | 0.04 | 0.15 | 0.29 | 2.81 | 0.39 | 0.14 | 0.54 | 0.06 | 0.03 |
| CH | $k_c$ | 0.02 | 2.18 | 2.52 | 4.43 | 0.10 | 0.10 | 14.48 | 1.20 | 0.02 |
| CH | n | 1.01 | 4.93 | 8.66 | 23.09 | 0.63 | 0.39 | 44.40 | 13.69 | 0.00 |
| TAN | $v_f$ | 0.04 | 0.18 | 0.04 | 3.13 | 0.08 | 0.04 | 1.52 | 0.12 | 0.03 |
| TAN | $k_j$ | 1.45 | 0.27 | 13.05 | 40.72 | 5.78 | 0.72 | 16.49 | 7.82 | 1.03 |
| TAN | n | 2.62 | 2.35 | 7.52 | 43.25 | 0.67 | 0.18 | 41.81 | 2.54 | 0.24 |

*The abbreviations are mentioned in Table 1

## 6 Conclusions and Way Forward

This study developed and validated a comprehensive Bayesian stochastic calibration framework for seventeen macroscopic traffic speed–density fundamental diagram models, integrating sequential Metropolis-within-Gibbs MCMC estimation with a novel Bias–Variance Cone analysis for asymptotic parameter identification. Applied to one full year of 5-minute PeMS loop detector data from three California freeway stations, the framework delivers the first formal application of asymptotic bias–variance decomposition theory to traffic FD parameter convergence, providing the first principled, quantitative, and empirically validated answers to the minimum data requirements for reliable FD parameter identification across a comprehensive model ensemble.

The seventeen models partition into three distinct classes based on posterior convergence behavior, parameter identifiability, and predictive accuracy. Nine models, GS, UW, NW, PM, DK, DR, CT, CH, and TAN, achieved stable posterior convergence with physically interpretable parameter estimates and RMSE values between 2.5 and 5 mph, demonstrating that model structure rather than estimation method is the primary determinant of Bayesian calibration reliability. Station 401671 produced the highest overall $R^2$ values across well-identified models, with CH achieving a median $R^2$ of 0.957, followed by TAN (0.954), Lee (0.951), and 5PL (0.950), while Station 401892 produced the lowest absolute $R^2$ values, a pattern attributable to differences in lane configuration and traffic heterogeneity rather than model structure. Critically, the standard deviations of $R^2$ across all 365 sampling iterations remained small for all well-identified models at all three stations, confirming that predictive performance is stable and not sensitive to the specific random subset of days selected for calibration. This stability implies that $R^2$ values achieved from as few as 30 to 50 days of data are reliable indicators of full-year performance.

Four models, GB, KN, NWL, and GD, produced catastrophic numerical failures attributable to logarithmic singularity at low density, unit system incompatibility, 1/k divergence under free-flow conditions, and extreme initialization sensitivity, respectively, and should not be applied to high-frequency loop detector data without fundamental structural modification. The remaining models, KH and PG, present a critical methodological warning: their competitive median $R^2$ values mask episodic catastrophic overflow events that standard single-metric reporting would entirely conceal, as demonstrated by KH achieving a median $R^2$ of 0.825 at Station 401892 while simultaneously producing a mean $R^2$ of $1.43 \times 10^{31}$. This finding underscores the necessity of examining the full $R^2$ distribution rather than central tendency alone when evaluating model reliability.

The Bias–Variance Cone framework established the cone-trajectory RMSE as a diagnostic of structural incompatibility, provided it is interpreted by parameter class rather than against a single absolute threshold. Free-flow speed parameters of well-identified models yielded cone RMSE below 1, whereas density parameters ($\boldsymbol{k_j}$, $\boldsymbol{k_c}$) yielded comparatively higher values of approximately 10–40, corresponding to normalized errors (NRMSE = cone RMSE/ $\theta_\infty$) of roughly 2–33%. This difference does not imply poor identification: density parameters are inherently harder to estimate because congested observations are under-represented relative to free-flow observations, which biases calibration (Qu et al. (2015); Cheng et al. (2022)), and because the all-sample-size cone RMSE is dominated by the wide, noisy early part of the trajectory, consistent with the longer data requirement identified below. In contrast, structurally incompatible models produced cone RMSE above 100 and NRMSE of several hundred percent, with NWL-$\boldsymbol{k_j}$ reaching 2221, KN-$\boldsymbol{k_c}$ reaching 2194, and GD-$\boldsymbol{k_c}$ reaching 1226, typically accompanied by opposite-sign bias coefficients and extreme deviations at 182 days (e.g., GB-$\boldsymbol{k_j}$, 1131% at Station 401657).

The sampling duration analysis yielded empirically derived minimum data thresholds for FD parameter identification: free-flow speed ($\boldsymbol{v_f}$) achieves reliable identification within approximately 30 days, with finite-sample bias consistently below 2%; jam density ($\boldsymbol{k_j}$) and critical density ($\boldsymbol{k_c}$) require 100 to 182 days for comparable stability, owing to the relative scarcity of congested observations in typical annual data cycles; and shape parameters in multi-parameter models exhibit persistent uncertainty beyond 182 days. Seven-day datasets produced deviations exceeding 500% for congestion-related parameters across multiple models and stations, confirming that one week of data is insufficient for reliable asymptotic FD parameter estimation regardless of model structure. These thresholds were broadly consistent across all three stations despite differences in lane configurations and congestion regimes, supporting their use as general guidelines for assessing FD calibration data sufficiency in freeway monitoring contexts.

The reliability of Bayesian FD calibration depends fundamentally on three factors: the structural compatibility of the model formulation with the prevailing data regime, the sufficiency of the observation period relative to parameter-specific data requirements, and the availability of congested-traffic observations for density parameter identification. Practical implications include minimum monitoring duration guidelines of 30 days for free-flow speed parameters, 100 to 182 days for density parameters, and 182 days or more for shape parameters in newly deployed detector stations; and model selection criteria that exclude GB, KN, NWL, and GD from high-frequency loop detector applications while treating KH and PG with caution unless full distributional diagnostics are conducted. Future work should extend the framework to heterogeneous urban datasets from developing countries, trajectory-based data from radar and video sensors, and formal Gelman-Rubin convergence diagnostics for highly parameterized models.